\documentclass[aps,amsfonts,pra,preprint,showpacs]{revtex4}
\usepackage{epsfig,amsmath,amssymb,bm,epsf,graphics,psfrag}

\def\ket#1{\vert#1\rangle}

\def\ipr#1#2{\langle#1\vert#2\rangle}

\def\Longarrow{\protect\@lra}
\def\@lra{\relbar\joinrel\relbar\joinrel\relbar\joinrel%
          \relbar\joinrel\rightarrow}

\begin{document}
\title{Global geometric entanglement in transverse-field XY spin chains: finite and infinite systems}

\author{Tzu-Chieh Wei}
\affiliation{Department of Physics and Astronomy, University of British
Columbia, Vancouver, BC V6T 1Z1, Canada}
\author{Smitha Vishveshwara}
\affiliation{Department of Physics and Institute for Condensed Matter Theory,
University of Illinois at Urbana-Champaign, Urbana, Illinois 61801, U.S.A.}
\author{Paul M. Goldbart}
\affiliation{Department of Physics, Institute for Condensed Matter Theory,
 and Federick Seitz Materials Research Laboratory, University of Illinois at Urbana-Champaign, Urbana,
Illinois 61801, U.S.A.}

\begin{abstract}
The entanglement in quantum XY spin chains of arbitrary length is investigated
via the geometric measure of entanglement. The emergence of entanglement is
explained intuitively from the perspective of perturbations. The model is
solved exactly and the energy spectrum is determined and analyzed in
particular for the lowest two levels for both finite and infinite systems.
The overlaps for these two levels are calculated analytically for arbitrary
number of spins. The entanglement is hence obtained by maximizing over a
single parameter. The corresponding ground-state entanglement surface is then
determined over the entire phase diagram, and its behavior can be used to
delineate the boundaries in the phase diagram.
For example, the field-derivative of the entanglement becomes singular along
the critical line. The form of the divergence is derived analytically and it
turns out to be dictated by the universality class controlling the quantum
phase transition. The behavior of the entanglement near criticality can be
understood via a scaling hypothesis, analogous to that for free energies. The
entanglement density vanishes along the so-called disorder line in the phase
diagram, the ground space is doubly degenerate and spanned by two product
states.
The entanglement for the superposition of the lowest two states is also
calculated. The exact value of the entanglement depends on the specific form
of superposition. However, in the thermodynamic limit the entanglement density
turns out to be independent of the superposition. This proves that the
entanglement density is insensitive to whether the ground state is chosen to
be the spontaneously $Z_2$ symmetry broken one or not. The finite-size scaling
of entanglement at critical points is also investigated from two different
view points. First, the maximum in the field-derivative of the entanglement
density is computed and fitted to a logarithmic dependence of the system size,
thereby deducing the correlation length exponent for the Ising class using
only the behavior of entanglement. Second, the entanglement density itself is
shown to possess a correction term inversely proportional to the system size,
with the coefficient being universal (but with different values for the ground
state and the first excited state, respectively).
\end{abstract}
 \maketitle

\section{Introduction}
Entanglement has been recognized in the past decade as a useful
resource in quantum information
processing~\cite{BennettBrassardCrepeauJozsaPeresWootters93,BennettWiesner92,Ekert91,NielsenChuang00}.
Very recently, it has emerged as an actor on the nearby stage of
quantum many-body physics, especially for systems that exhibit
quantum phase
transitions~\cite{manybody,OsborneNielsen02,OsterlohAmicoFalciFazio02,VidalLatorreRicoKitaev03,CalabreseCardy04,Korepin,SommaOrtizBarnumKnillViola04},
where it can play the role of a diagnostic of quantum correlations.
Quantum phase transitions~\cite{Sachdev} are transitions between
qualitatively distinct phases of quantum many-body systems, driven
by quantum fluctuations. It is conventionally defined as the
non-analyticity in the ground-state energy~\cite{Sachdev}.  In view
of the connection between entanglement and quantum correlations, one
anticipates that entanglement will furnish a dramatic signature of
the quantum critical point as the ground-state wavefunction itself
should also exhibit certain non-analyticity, which thereby manifests
in the singularity of correlations. To describe the ground state
usually requires exponentially many parameters, and therefore to
detect the non-analyticity hidden in the wavefunction may require an
ingenious design of observables. Ground-state energy is usually such
a good indicator but it may fail to exhibit singularity, for
example, near the isotropic antiferromagnetic point in the XXZ
model~\cite{YangYang}, where there is a continuous transition.
Entanglement is thus expected to be an alternative, useful quantity
in assisting the determination of quantum phase transitions,
complementing traditional methods. On the other hand, the more
entangled a state is, usually, though not necessarily, the more
useful it is likely to be a resource for quantum information
processing~\cite{TooMuch}.
 It is thus desirable to study and quantify
the degree of entanglement near quantum phase transitions.  By employing
entanglement to diagnose many-body quantum states one may obtain fresh insight
into the quantum many-body problem~\cite{manybody}.

To date, progress in characterizing and quantifying entanglement has
taken place greatly in the domain of bi-partite systems, and much
understanding has be gained in the multi-partite settings, although
a complete picture has not yet been achieved~\cite{Entanglement}.
Much of the previous work on entanglement in quantum phase
transitions has been based on bi-partite measures, i.e., focus has
been on entanglement either between pairs of
parties~\cite{OsborneNielsen02,OsterlohAmicoFalciFazio02,Roscilde04}
or between a part and the remainder of a
system~\cite{VidalLatorreRicoKitaev03,CalabreseCardy04,Korepin,ChungPeschel,EEmore}.
For multi-partite systems, however, the complete characterization of
entanglement requires the consideration of multi-partite
entanglement, for which a consensus measure has not yet emerged.

Singular and scaling behavior of entanglement near quantum critical points was
discovered in important work by Osterloh and
co-workers~\cite{OsterlohAmicoFalciFazio02}, who invoked Wootters' {\it
bi-partite\/} concurrence~\cite{Wootters98} as a measure of entanglement. The
success of the use of concurrence has been understood via its connection to
two-point correlation functions and the derivative of ground-state
energy~\cite{two-reduced}. The drawback of concurrence is that it can deal
with only two spins (each with spin-1/2) even though the system may contain an
infinite number of spins. Although attempts have been made to generalize
concurrence to many spin-1/2 systems via the time reversal operation, the
generalized concurrence loses its connection to the entanglement of
formation~\cite{WongChristensen01}. Furthermore, the use of concurrence may
give rise to spurious transition points~\cite{Yang}.

Another approach is to consider the von Neumann entropy of a
subsystem of $L$ spins with the rest $N-L$ spins of the system. It
is found that for critical spin chains the entropy scales
logarithmically with the subsystem size $L$ for
$N\rightarrow\infty$, with a prefactor that is related to the
central charge of the corresponding conformal
theory~\cite{VidalLatorreRicoKitaev03,CalabreseCardy04,Korepin}. The
entanglement addressed in this case is  only between a subsystem and
the rest of the system, and the connection to central charge is
interesting.

A different approach without using entanglement is to consider fidelity or
overlap between two different ground-state wavefunctions, which is now called
fidelity measure. Non-analytic behavior in the fidelity measure is expected
for system parameters close to critical points. This approach has been shown
to be successful~\cite{fidelity}.

Recently, Barnum and co-workers~\cite{SommaOrtizBarnumKnillViola04} have
developed an entanglement measure, which they call generalized entanglement.
Instead of using subsystems they use different algebras and generalized
coherent states to define the entanglement. They have also applied the
generalized entanglement to systems exhibiting quantum phase transitions.
Their approach opens a new approach to multi-partite entanglement.
Nevertheless, there is no {\it a priori\/} choice of which algebra, amongst
all possible ones, is the most natural one to use.

In addition to spin chains other models that have been studied by using either
the von Neumann entropy (also known as entanglement entropy) or the
concurrence as the entanglement measure include: (i) the super-radiance model,
in which many two-level atoms interact with a single-mode photon
field~\cite{LambertEmaryBrandes04}; and (ii) the one-dimensional extended
Hubbard model, in which electrons can hop between the nearest neighbors and
there are Coulomb interactions among electrons on the same site and with
nearest-neighbor electrons as well~\cite{GuDengLiLin04}. Furthermore, in
addition to spin systems, entanglement entropy has been examined in both
bosonic and fermionic systems~\cite{EEmore,Fermion} and the area law (or the
breaking of it) has been investigated. In view of other approaches that
connect quantum information ideas to condensed matter settings (or vice
versa), Verstraete and co-workers~\cite{VerstraeteMartin-DelgadoCirac04} have
 defined an entanglement length, viz., the distance at which two sites can
establish a pure-state entanglement at the cost of measuring all other sites.
They found that this entanglement length is lower bounded the correlation
length. All these, including the theme of the present paper, are aimed at
approaching many-body problems from different, and hopefully fresh,
perspectives, and at complementing traditional statistical-physical
approaches.

In the present paper, we expand and extend results of Ref.~\cite{WeiEtAl05}.
We apply the {\it global\/} geometric measure that we have developed in
Ref.~\cite{WeiGoldbart03}, based on a geometric
picture~\cite{Shimony,BarnumLinden01,Arecchi72,Brody}; it provides a {\it
holistic\/}, rather than bi-partite, characterization of the entanglement of
quantum many-body systems. Our focus is on one-dimensional spin systems,
specifically ones that are exactly solvable and exhibit quantum criticality.
The geometric measure has been applied to many models in one
dimension~\cite{OtherGE,OrusWei,blockGE,RGGE,Son} as well as two
dimensions~\cite{huang} and seems to become a useful tool as well. For these
systems,  it is found that entanglement behaves in a singular manner near the
quantum critical points. This supports the view that entanglement---the
non-factorization of wave functions---reflects quantum correlations. This will
be explained in detail using the transverse-field XY models as an specific
example.

The remaining structure of the present paper is as follows. In
Sec.~\ref{sec:GE}, we introduce the geometric measure of entanglement (or in
short, geometric entanglement) that we shall use to quantify the entanglement
of a many-body state. In Sec.~\ref{sec:XY}, we introduce the spin XY models in
a transverse field, the focus of present paper.  We explain in an elementary
way the emergence of entanglement away from critical points, so as to expect
the drastic behavior of entanglement near criticality. In Sec.~\ref{sec:Diag},
we diagonalize the Hamiltonian, invoking Jordan-Wigner and Bogoliubov
transformations. We carefully explain the boundary conditions placed on the
spin and hence the fermionic operators. In Sec.~\ref{sec:overlap}, we derive
the overlap of the lowest two levels with an Ansatz product state, the result
of which is the basis for deriving entanglement for these two states,
including the true ground state. Section~\ref{sec:EntMain} contains the main
results of the present paper. We present the results of ground-state
entanglement over the entire phase diagram. In Sec.~\ref{subsec:finite}, we
illustrate the entanglement results for finite systems. In
Sec.~\ref{subsec:infinite}, we derive the expression of entanglement for
infinite systems. There, the divergence of ground-state entanglement is
analyzed and the connection to the universality classes is established. In
Sec.~\ref{subsec:disorder}, we show the vanishing of entanglement along the
disorder line by demonstrating the ground space is doubly degenerate and
spanned by two product states. In Sec.~\ref{subsec:superposition}, we discuss
the issue of entanglement superposition, and argue the irrelevance of
spontaneous symmetry breaking in the determination of ground-state
entanglement density. In Sec.~\ref{subsec:finite-size}, we study the issue of
finite-size effect to the entanglement density at criticality. Finally, we
summarize in Sec.~\ref{sec:conclude}, and give a generic picture of how
entanglement behaves near criticality, using a scaling hypothesis. How the
geometric entanglement serves to detect non-analyticity of ground-state
wavefunction is also discussed.

\section{Global geometric measure of entanglement}
\label{sec:GE} We quickly review the global measure that we shall use in the
present paper. Consider a general, $n$-partite, normalized pure state:
$|\Psi\rangle=\sum_{p_1\cdots p_n}\Psi_{p_1p_2\cdots p_n}
|e_{p_1}^{(1)}e_{p_2}^{(2)}\cdots e_{p_n}^{(n)}\rangle$. If the parties are
all spin-1/2 then each can be taken to have the basis
$\{\ket{\!\uparrow},\ket{\!\downarrow}\}$. Our scheme for analyzing the
entanglement involves considering how well an entangled state can be
approximated by some unentangled (normalized) state (e.g.,~the state in which
every spin points in a definite direction):
$\ket{\Phi}\equiv\mathop{\otimes}_{i=1}^n|\phi^{(i)}\rangle$. The proximity of
$\ket{\Psi}$ to $\ket{\Phi}$ is captured by their overlap; the entanglement of
$\ket{\Psi}$ is revealed by the maximal
overlap~\cite{Shimony,BarnumLinden01,Biham,WeiGoldbart03,Arecchi72,Brody}
\begin{equation}
\label{eq:lambdamax}
\Lambda_{\max}({\Psi})\equiv\max_{\Phi}|\ipr{\Phi}{\Psi}|\,;
\end{equation}
the larger $\Lambda_{\max}$ is, the less entangled is $\ket{\Psi}$. (Note that
for a product state, $\Lambda_{\max}$ is unity.) If the entangled state
consists of two separate entangled pairs of subsystems, $\Lambda_{\max}$ is
the product of the maximal overlaps of the two. Hence, it makes sense to
quantify the entanglement of $\ket{\Psi}$ via the following {\it extensive\/}
quantity~\cite{foot:REE}
\begin{equation}
E_{\log_2}({\Psi})\equiv-\log_2\Lambda^2_{\max}(\Psi),
\label{eq:Entrelate}
\end{equation}
This normalizes to unity the entanglement of EPR-Bell and $N$-party GHZ states,
as well as giving zero for unentangled states.
Finite-$N$ entanglement is interesting in the context of
quantum information processing. To characterize the properties of the
quantum critical point we use the thermodynamic quantity ${\cal E}$ defined
by
\begin{subequations}
\begin{eqnarray}
&&{\cal E}\equiv\lim_{N\to\infty}{\cal E}_{N}, \\
&&{\cal E}_{N}\equiv
{N}^{-1}E_{\log_2}(\Psi),
\end{eqnarray}
\end{subequations}
where ${\cal E}_{N}$ is the {\it entanglement density\/}, i.e., the entanglement per particle.

For translation invariant state, it is shown that the entanglement is either
completely in a product state or is globally entanglement~\cite{symmetry}.
Therefore, the above quantification of entanglement by comparing to product
states makes a natural choice.

We remark that by appropriate partitioning of $\ket{\phi}$ into
various product forms, a hierarchy of entanglement can be
obtained~\cite{ShimoniBiham07,BlasoneDellAnnoDeSienaIlluminati08}.
This hierarchical geometric measure has recently also been applied
to the XY
model~\cite{BlasoneDellAnnoDeSienaIlluminati09,BlasoneDellAnnoDeSienaIlluminati10}.
One can generalize the form of product states, e.g., to a product of
blocks of two spins (or multiple spins), and then calculate the
entanglement per block~\cite{blockGE}. In fact, this approach
enables the connection of the spatial renormalization group (RG)
(i.e., coarse-graining) procedure to the geometric measure, and the
entanglement under RG can be properly defined~\cite{RGGE}.
\section{Quantum XY spin chains and emergence of entanglement}
\label{sec:XY}
 We consider the family of models governed by the Hamiltonian
\begin{equation}
\label{eqn:HXY}
{{\cal H}_{\rm XY}}=
- \sum_{j=1}^N \left(\frac{1\!+\!r}{2}
\sigma_j^x\sigma_{j\!+\!1}^x+
\frac{1\!-\!r}{2}\sigma_j^y\sigma_{j\!+\!1}^y+
h \,\sigma_j^z\right),
\end{equation}
where $r$ measures the anisotropy between $x$ and $y$ couplings,
$h$ is the transverse external field, lying along the $z$-direction,
and we impose periodic boundary conditions, namely, a ring geometry.
At $r=0$ we have the isotropic XY limit (also known as the XX model)
and
at $r=1$, the Ising limit. All anisotropic XY models
($0<r\le 1$) belong to the same universality class, i.e., the Ising
class, whereas the isotropic XX model belongs to a different universality
class.
XY models exhibit three phases (see Fig.~\ref{fig:XYEnt10000}):
oscillatory, ferromagnetic and paramagnetic.
In contrast to the paramagnetic phase, the first two are ordered phases,
with the oscillatory phase being associated
with a characteristic wavevector, reflecting the modulation
of the spin correlation functions (see, e.g., Ref.~\cite{Henkel99}).
We shall see that the global
entanglement detects the boundaries between these phases, and that
the universality class dictates the behavior of entanglement
near quantum phase transitions.

Before we solve the entanglement of the XY model, we give perturbative
analysis of, as an
illustration of how entanglement arises and vanishes, the Ising model in a transverse field (viz. $r=1$)
\begin{equation}
\label{eqn:Hising}
{\cal H}=- \sum_{i=1}^N \left(\sigma_i^x\sigma_{i+1}^x+h \sigma_i^z\right).
\end{equation}
At $h=0$ the ground state is that with all spins pointing up in the $x$-direction $\ket{\!\!\rightarrow\rightarrow\dots\rightarrow}$ or down $\ket{\!\!\leftarrow\leftarrow\dots\leftarrow}$,
which is manifestly unentangled.
The ground state can be any superposition of $(\ket{\!\!\rightarrow\rightarrow\dots\rightarrow}$ and  $\ket{\!\!\leftarrow\leftarrow\dots\leftarrow}$ when the $Z_2$ symmetry is not
spontaneously broken. For example, the states
$(\ket{\!\!\rightarrow\rightarrow\dots\rightarrow}\pm\ket{\!\!\leftarrow\leftarrow\dots\leftarrow})/\sqrt{2}$ are actually the two lowest levels obtained
from solving the models using Jordan-Wigner and Bogoliubov transformations
and they both have $E_{\log_2}=1$. (For small $h$ the entanglement rises quadratically in the case of unbroken symmetry instead of quartically, as
we shall show shortly.) We shall see later that whether or not we use a broken-symmetry
state  has no effect in the thermodynamic limit.
 For small $h$ (i.e., $h\ll1/\sqrt{N}$) one can
obtain the ground state by treating the $h \sigma_i^z$ terms as perturbations.
Take the ground state
at $h=0$ to be $\ket{\!\!\rightarrow\rightarrow\dots\rightarrow}$. Then first-order perturbation theory for the ground state gives
\begin{equation}
\frac{1}{\sqrt{1+\frac{Nh^2}{16}}}\Big(\ket{\!\!\rightarrow\rightarrow\dots\rightarrow}+\frac{h}{4}\sum_i\ket{\!\!\rightarrow\dots\leftarrow_{i}\dots\rightarrow}\Big).
\end{equation}
Using the method described in Ref.~\cite{WeiGoldbart03} we obtain
$E_{\log_2}\lesssim N h^2/(16\ln2)$ to leading order in $h$.
 At $h=\infty$ the ground state is a quantum paramagnet with all
spins aligning along the external field: $\ket{\!\!\uparrow\uparrow\dots\uparrow}$, and once more is
 unentangled.
To ${\cal O}(1/h)$ perturbation theory gives (treating $\sigma^z_i\sigma^z_{i+1}$ terms small)
\begin{equation}
\frac{1}{\sqrt{1+\frac{N}{16h^2}}}\Big(\ket{\!\!\uparrow\uparrow\dots\uparrow}+\frac{1}{4h}\sum_i\ket{\!\!\uparrow\dots\downarrow_{i}\downarrow_{{i\!+\!1}}\dots\uparrow}\Big),
\end{equation}
for which $E_{\log_2}\lesssim N/(16 h^2\ln2)$, to leading order of
$1/h$. The quantum phase transition from a ferromagnetic to a
paramagnetic phase occurs at $h=1$~\cite{Sachdev}. The two lowest
levels, which we denote by $\ket{\Psi_{1/2}}$ and $\ket{\Psi_0}$
(for reasons to be explained later) are, respectively, the ground
and first excited states, and they are asymptotically degenerate for
$0\le h\le1$ when $N\rightarrow\infty$.

\section{Diagonalization of the Hamiltonian}
\label{sec:Diag}
 As is well known~\cite{Sachdev,Henkel99,LiebSchultzMattis61},
the energy eigenproblem for the XY spin chain can be solved via a
Jordan-Wigner transformation, through which the spin degrees of freedom are
recast as fermionic ones, followed by a Bogoliubov transformation, which
diagonalizes the resulting quadratic Hamiltonian.

The Jordan-Wigner transformation that we shall make from
spins ($\sigma$'s) to fermion particles ($c$'s) is
\begin{subequations}
\begin{eqnarray}
\sigma_i^z&=&1-2c_i^\dagger\,c_i, \\
\sigma_i^x&=&\prod_{j=1}^{i-1}\big(1-2c_j^\dagger\,c_j\big)\big(c_i+c_i^\dagger\big),\\
\sigma_i^y&=&-i\prod_{j=1}^{i-1}\big(1-2c_j^\dagger\,c_j\big)\big(c_i-c_i^\dagger\big).
\end{eqnarray}
\end{subequations}
One has to pay attention to the boundary conditions that are to be
imposed on the $c$'s. Although
periodic in the $\sigma$'s, one cannot simply
take
\begin{equation}
\sigma_{N+1}^x=\prod_{j=1}^{N}\big(1-2c_j^\dagger\,c_j\big)
\big(c_{N+1}+c_{N+1}^\dagger\big)=\sigma_1=(c_{1}+c_{1}^\dagger\big),
\end{equation}
and conclude that either (i)~$\prod_{j=1}^{N}\big(1-2c_j^\dagger\,c_j\big)=1$
and $c_{N+1}=c_1$, or (ii)~$\prod_{j=1}^{N}\big(1-2c_j^\dagger\,c_j\big)=-1$
and $c_{N+1}=-c_1$, neither of which are correct if one wishes to obtain
the correct spectrum and eigenstates for arbitrary finite $N$. Instead one should  impose that (as the $(N+1)$-th site is identified as the first site)
\begin{equation}
\sigma_{N}^x\sigma_{N+1}^x=\sigma_{N}^x\sigma_{1}^x,
\end{equation}
which then leads to
\begin{equation}
\label{eqn:bc}
\big(c_{N}+c_{N}^\dagger\big)\big(c_{N+1}+c_{N+1}^\dagger\big)
=-\prod_{j=1}^{N}\big(1-2c_j^\dagger\,c_j\big)\big(c_{N}+c_{N}^\dagger\big)\big(c_{1}+c_{1}^\dagger\big).
\end{equation}
The two possible conditions that satisfy this are
either (I)~$\prod_{j=1}^{N}\big(1-2c_j^\dagger\,c_j\big)=-1$
and $c_{N+1}=c_1$, or (II)~$\prod_{j=1}^{N}\big(1-2c_j^\dagger\,c_j\big)=1$
and $c_{N+1}=-c_1$.
The operator
\begin{equation}
\prod_{j=1}^{N}\big(1-2c_j^\dagger\,c_j\big)=e^{i\pi\sum_j c_j^\dagger\,c_j}
\end{equation}
counts whether the total number of particles is even ($+1$) or odd ($-1$).
For $c$'s that are periodic, the number is odd, whereas for antiperiodic $c$'s,
this number is even.

To incorporate these two boundary conditions on the $c$'s, we take
\begin{equation}
c_j=\frac{1}{\sqrt{N}}\sum_{m=0}^{N-1}e^{i\frac{2\pi}{N}j (m+b)} \tilde{c}_{m}^{(b)},
\end{equation}
where $b=0$ for periodic $c$'s; $b=1/2$ for anti-periodic $c$'s.
(This explains why we label the lowest two states by $\ket{\Psi_{1/2}}$ and $\ket{\Psi_0}$.)
The momentum index $m$ ranges from $0$ to $N-1$.
In terms of these fermion operators the Hamiltonian becomes
\begin{eqnarray}
{\cal H}&=&-N h
-\sum_{m=0}^{N-1}\left\{\Big[2\cos\frac{2\pi}{N}(m+b)-2h\Big]{\tilde{c}_{m}^{(b)\dagger}}
\tilde{c}_{m}^{(b)}\right.\nonumber\\
&&\left. +i r \sin\frac{2\pi}{N}(m+b)\Big[{\tilde{c}_{m}^{(b)}}
\tilde{c}_{N-m-2b}^{(b)}+{\tilde{c}_{m}^{(b)\dagger}}
{\tilde{c}_{N-m-2b}^{(b)\dagger}}\Big]\right\}.
\end{eqnarray}

Upon using the Bogoliubov transformation
\begin{equation}
\tilde{c}_m^{(b)}=\cos\theta_m^{(b)} \gamma_m^{(b)}+i \sin\theta_m^{(b)} \gamma^{(b)\dagger}_{N-m-2b}\,,
\end{equation}
with
\begin{equation}
\tan2\theta_m^{(b)}={r\sin\frac{2\pi\,(m+b)}{N}}\Big/{\left(h-\cos\frac{2\pi\,(m+b)}{N}\right)},
\end{equation}
one arrives at the diagonal the Hamiltonian:
\begin{subequations}
\begin{equation}
{\cal
H}=-Nh+\sum_{m=0}^{N-1}\varepsilon_m^{(b)}\Big(\tilde{\gamma}_{m}^{(b)\dagger}
\tilde{\gamma}_{m}^{(b)}-\frac{1}{2}\Big),
\end{equation}
\begin{equation}
\varepsilon_m^{(b)}=2\sqrt{\Big(h-\cos\frac{2\pi\,(m+b)}{N}\Big)^2+r^2\sin^2\frac{2\pi\,(m+b)}{N}},
\end{equation}
\end{subequations}
except for the special case that $\varepsilon_0^{(0)}=2(h-1)$.

\begin{figure}
\centerline{\psfig{figure=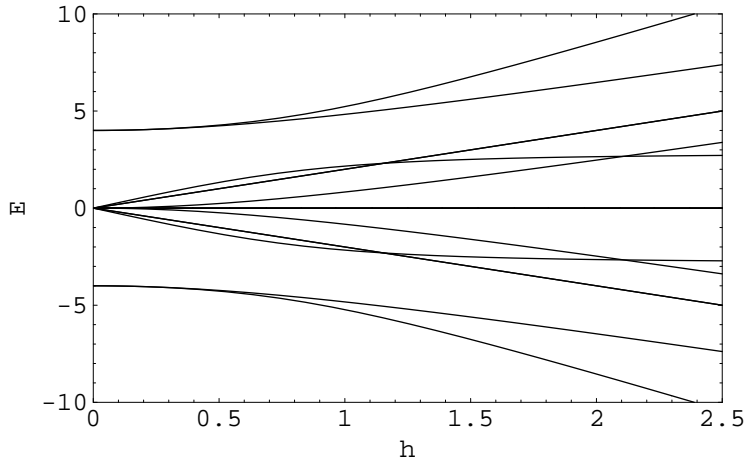,width=10cm,angle=0}}
\caption{Energy spectrum vs.~magnetic field $h$ for the Ising spin chain
with $N=4$ qubits from the Bogoliubov diagonalization. The results are identical to exact matrix diagonalization.   }
\label{fig:E_Ising4}
\end{figure}
\begin{figure}
\centerline{\psfig{figure=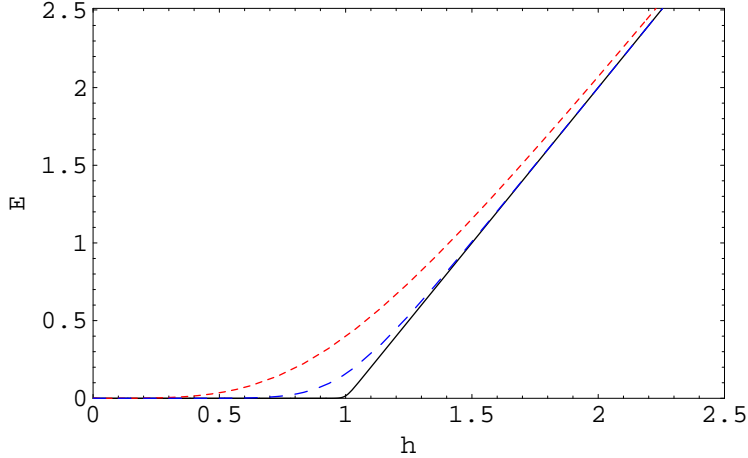,width=10cm,angle=0}} \caption{(Color
online) Energy gap vs.~magnetic field $h$ for the Ising spin chain with $N=4$
(red shorter dashed), $N=10$ (blue longer dashed), and $N=100$ (black solid
curve).   } \label{fig:Gap_Ising}
\end{figure}
We remark that we have not left out any constant in diagonalizing the
Hamiltonian in either case, so the energy spectrum is exact.
For each value of $b$ the diagonalization gives $2^N$ energy eigenvalues,
so there
are $2^{N+1}$ in total. Half of them are spurious. In determining the
correct $2^N$ states from the $2^{N+1}$ solutions, one has to impose
a constraint from the boundary conditions. Namely, in case I there
can be only odd number of fermions, whereas in case II
there can be only even number of fermions.

For $b=0$, viz, the odd-number-fermion case, the lowest state $\ket{\Psi_{0}}$
is such that $\langle \tilde{\gamma}_{m}^{(0)\dagger} \tilde{\gamma}_{m}^{(0)}\rangle=0$ except that $\langle \tilde{\gamma}_{0}^{(0)\dagger} \tilde{\gamma}_{0}^{(0)}\rangle=1$.
Its energy eigenvalue is
\begin{equation}
\label{eqn:E0}
E_0^{(0)}(r,h)=(h-1)-\sum_{m=1}^{N-1}\sqrt{\Big(h-\cos\frac{2\pi\,m}{N}\Big)^2+r^2\sin^2\frac{2\pi\,m}{N}}.
\end{equation}

For $b=1/2$, namely, the even-number-fermion case, the lowest state $\ket{\Psi_{1/2}}$
is such that $\langle \tilde{\gamma}_{m}^{(1/2)\dagger} \tilde{\gamma}_{m}^{(1/2)}\rangle=0$ for all $m$.
Its eigen-energy is
\begin{equation}
\label{eqn:E1/2}
E_0^{(1/2)}(r,h)=-\sum_{m=0}^{N-1}\sqrt{\Big(h-\cos\frac{2\pi\,(m+1/2)}{N}\Big)^2+r^2\sin^2\frac{2\pi\,(m+1/2)}{N}}.
\end{equation}

We see that, as $N\rightarrow \infty$, the above two energy levels are degenerate
for $h\le 1$. Furthermore,  as $N\rightarrow\infty$ the difference between the two energy levels becomes
\begin{equation}
\label{eqn:gap}
E_0^{(0)}(r,h)-E_0^{(1/2)}(r,h)=2(h-1) \Theta(h-1),
\end{equation}
where $\Theta(x)=1$ if $x>0$ and zero otherwise. The way the energy gap
vanishes as $h\rightarrow 1^+$ gives a relation between two exponents
\begin{equation}
\label{eqn:znu}
z\nu=1;
\end{equation}
$z$ is the dynamical exponent (defined via the vanishing of energy gap
$\Delta\sim|h-h_c|^{z\nu}$) and $\nu$ is the correlation-length
exponent (defined via $L_c\sim |h-h_c|^{-\nu}$).

\begin{figure}
\centerline{\psfig{figure=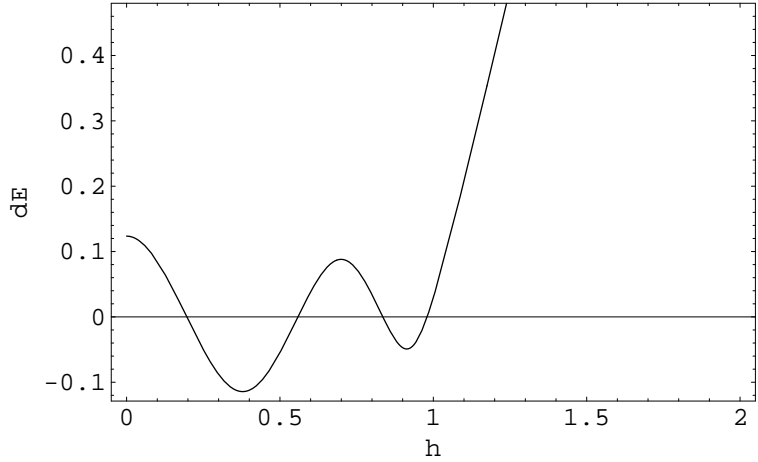,width=10cm,angle=0}} \caption{Energy Energy
difference $dE\equiv E^{(0)}_0-E^{(1/2)}_0$ vs.~magnetic field $h$ for the XY
spin chain with $r=0.2$ and $N=8$ qubits from the Bogoliubov diagonalization.
  } \label{fig:gxDE}
\end{figure}

We show in Fig.~\ref{fig:E_Ising4} the energy spectrum for $N=4$ Ising spin
chain obtained from the above JW and Bogoliubov diagonalization, which is
identical to direct matrix diagonalization (not shown). This means that the
choice of our boundary conditions below Eq.~(\ref{eqn:bc}) is correct. We
illustrate the gap in the spectrum for the Ising chain in
Fig.~\ref{fig:Gap_Ising}, which is obviously degenerate for $h<1$ and linear
($h-1$) for $h>1$, as derived in Eq.~(\ref{eqn:gap}). We see that $E^{(0)}_0$
is asymptotically (i.e., as $N\rightarrow\infty$) degenerate with
$E^{(1/2)}_0$ in $h\in[0,1]$, but the former has higher energy than the latter
for $h>1$. For finite $N$, $E^{(0)}_0$ is larger than $E_{0}^{(1/2)}$.
However, this is not the case for general XY model, as we illustrate in
Fig.~\ref{fig:gxDE} for $r=0.2$, where we see that the two branches switch
roles of being the true ground state, depending on the value of $h$.

Even though the ground state may switch back and forth between
$\ket{\Psi_{1/2}}$ and $\ket{\Psi_{0}}$ for $h<1$, for $h\ge 1$
$\ket{\Psi_{1/2}}$ is the ground state. In particular, at $h=1$ we calculate
$\Delta_N(r)\equiv E_{0}^{(0)}(r,h=1)-E_{0}^{(1/2)}(r,h=1)$ for the finite
chain with $N$ spins and find that
\begin{equation}
\Delta_N(r)=\frac{\pi\, r}{2N} + {\cal O}(1/N^2).
\end{equation}
It is interesting to see that the difference in the two energies is
proportional to the anisotropy $r$ and, as expected, inversely proportional to
the system size $N$.

\section{Derivation of overlap of the lowest two states with the product Ansatz
state} \label{sec:overlap}
 Having found the lowest two eigenstates, the
quantity $\Lambda_{\rm max}$ of Eq.~(\ref{eq:lambdamax})---and hence the
entanglement---can be found, at least in principle. To do this, we parametrize
the separable states via
\begin{equation}
\ket{\Phi}\equiv\mathop{\otimes}_{i=1}^{N}
\big[\cos(\xi_i/2)\ket{\!\!\uparrow}_i +
e^{i\phi_i}\sin(\xi_i/2)\ket{\!\!\downarrow}_i\big],
\end{equation}
where $\ket{\!\uparrow\!\!/\!\!\downarrow}$ denote spin states parallel/antiparallel
to the $z$-axis.  Instead of maximizing the overlap with respect to the $2N$ real
parameters $\{\xi_{i},\phi_{i}\}$, for the lowest two states it is
adequate to appeal to the translational symmetry of and the reality of the
ground-state wavefunctions.  Thus taking $\xi_i=\xi$ and $\phi_i=0$ we make the Ansatz:
\begin{equation}
\label{eqn:PhiTh}
 \ket{\Phi(\xi)}
\equiv
e^{-i\frac{\xi}{2}\sum_{j=1}^N\sigma_j^y}
\ket{\!\uparrow\uparrow\dots\uparrow}
\end{equation}
for searching for the maximal the overlap $\Lambda_{\max}(\Psi)$~\cite{footnote:ansatz}.
This form shows that this separable state can be constructed
as a global rotation of the ground state at $h=\infty$, viz., the separable
state $\ket{\!\uparrow\uparrow\dots\uparrow}$.
In this particular limit the boundary condition on the $c$'s is irrelevant,
as the dominant term in the Hamiltonian is  $\,-\sum_j h(1-2c_j^\dagger c_j)$.

The energy eigenstates are readily expressed in terms of the Jordan-Wigner
fermion operators, and so too is the  family of the Ansatz states
$\ket{\Phi(\xi)}$.  By working in this fermion basis we are able to evaluate
the overlaps between the two lowest states and the Ansatz states, presented in
the next section.

  We first analyze $b=1/2$ case, viz., the even-fermion case. The lowest
state $\ket{\Psi_{1/2}(r,h)}$ has zero number of Bogoliubov fermions. It is
related to the state that has no $c$-fermions, i.e.,
$\ket{\Omega}\equiv\ket{\!\uparrow\cdots\uparrow}$ via
\begin{subequations}
\begin{eqnarray}
\ket{\Psi_{1/2}(r,h)}&=&\prod_{m=0}^{m<\frac{N\!-\!1}{2}}\cos\theta_m^{(1\!/2)}(r,h)
e^{i\tan\theta_m^{(1\!/2)}(r,h)\,\tilde{c}_m^{(1\!/2)\dagger}
\tilde{c}_{N-m-1}^{(1\!/2)\dagger} }\ket{\Omega}\\
&=&\prod_{m=0}^{m<\frac{N\!-\!1}{2}}\Big[\cos\theta_m(r,h)
+i\sin\theta_m(r,h)\,\tilde{c}_m^{\dagger} \tilde{c}_{N-m-1}^{\dagger}
\Big]\ket{\Omega}.
\end{eqnarray}
\end{subequations}
The Ansatz state is then
\begin{subequations}
\begin{eqnarray}
\ket{\Phi(\xi)} &=&\prod_{j=1}^N
\Big[ \cos\frac{\xi}{2}+\sin\frac{\xi}{2}\prod_{1\le l<j}(1-2c_l^\dagger c_l) (c_j^\dagger-c_j) \Big] \ket{\Omega} \\
&=&\prod_{j=1}^N\big(\cos\frac{\xi}{2}+\sin\frac{\xi}{2}\,c_j^\dagger\big)
\ket{\Omega}\\
&=&\cos^{N}\frac{\xi}{2}\, e^{\tan\frac{\xi}{2}c_1^\dagger}\cdots e^{\tan\frac{\xi}{2}c_N^\dagger}\ket{\Omega}\\
&=&\cos^{N}\frac{\xi}{2}\, e^{\tan\frac{\xi}{2}\sum_{j=1}^N c_j^\dagger}
e^{\tan^2\frac{\xi}{2}\sum_{j< l}c_j^\dagger c_l^\dagger}\ket{\Omega},
\end{eqnarray}
\end{subequations}
where we have suppressed the index $(1/2)$. The term $\sum_{j< l}c_j^\dagger
c_l^\dagger$
 can be rewritten in momentum space as
 \begin{equation}
 \sum_{1\le j< l\le N} c_j^\dagger c_l^\dagger=i\sum_{m=0}^{m<\frac{N-1}{2}}
 \cot\frac{\pi(m+\frac{1}{2})}{N}\tilde{c}^\dagger_m \tilde{c}^\dagger_{N\!-\!m\!-\!1} .
 \end{equation}
  Thus, for even $N$
\begin{subequations}
 \begin{equation}
 \ket{\Phi(\xi)}=\Big({1+\tan\frac{\xi}{2}\,\sum_{j=1}^N c_j^\dagger}\Big)\prod_{m=0}^{m<\frac{N-1}{2}}
\left(\cos^2\frac{\xi}{2}+i\sin^2\frac{\xi}{2}\cot\frac{\pi(m+\frac{1}{2})}{N}
\tilde{c}^\dagger_m \tilde{c}^\dagger_{N\!-\!m\!-\!1}\right)\ket{\Omega},
 \end{equation}
 whereas for odd $N$
 \begin{equation}
\ket{\Phi(\xi)}=\Big({1+\tan\frac{\xi}{2}\,\sum_{j=1}^N c_j^\dagger}\Big)
\cos\frac{\xi}{2}\prod_{m=0}^{m<\frac{N-1}{2}}
\left(\cos^2\frac{\xi}{2}+i\sin^2\frac{\xi}{2}\cot\frac{\pi(m+\frac{1}{2})}{N}
\tilde{c}^\dagger_m \tilde{c}^\dagger_{N\!-\!m\!-\!1}\right)\ket{\Omega}.
 \end{equation}
 \end{subequations}
 Therefore, the overlap of the state $\ket{\Psi_{1/2}(r,h)}$ with $\ket{\Phi(\xi)}$ for even $N$ is
 \begin{subequations}
 \begin{equation}
 \ipr{\Psi_{1/2}(r,h)}{\Phi(\xi)}=\prod_{m=0}^{m<\frac{N-1}{2}}
\left(\cos\theta_m^{(1/2)}(r,h)\,\cos^2\frac{\xi}{2}+\sin\theta_m^{(1/2)}(r,h)\,\sin^2\frac{\xi}{2}\cot\frac{\pi(m+\frac{1}{2})}{N}\right),
 \end{equation}
whereas  for odd $N$
 \begin{equation}
 \ipr{\Psi_{1/2}(r,h)}{\Phi(\xi)}=\cos\frac{\xi}{2}\prod_{m=0}^{m<\frac{N-1}{2}}
\left(\cos\theta_m^{(1/2)}(r,h)\,\cos^2\frac{\xi}{2}+\sin\theta_m^{(1/2)}(r,h)\,\sin^2\frac{\xi}{2}\cot\frac{\pi(m+\frac{1}{2})}{N}\right).
 \end{equation}
 \end{subequations}

 Next, we discuss the $b=0$ (odd-fermion) case. The lowest allowed state is
the one with one $\gamma^{(0)}_0=\tilde{c}^{(0)}_0$ fermion:
\begin{equation}
\ket{\Psi_{0}(r,h)}\equiv\gamma^{(0)\dagger}_0\ket{G(r,h)}=\tilde{c}^{(0)\dagger}_0\ket{G(r,h)},
\end{equation}
where $\ket{G(r,h)}$ is the state with no $\gamma$ fermions:
\begin{equation}
\ket{G(r,h)}=\prod_{m=1}^{m<\frac{N}{2}}\Big[\cos\theta_m^{(0)}(r,h)
+i\sin\theta_m^{(0)}(r,h)\,\tilde{c}_m^{(0)\dagger}
\tilde{c}_{N-m}^{(0)\dagger} \Big]\ket{\Omega}.
\end{equation}
Similar to the $b=1/2$ case, by using
\begin{equation}
 \sum_{1\le j< l\le N} c_j^\dagger c_l^\dagger=i\sum_{m=1}^{m<\frac{N}{2}}
 \cot\frac{\pi m}{N}\tilde{c}^{(0)\dagger}_m \tilde{c}^{(0)\dagger}_{N\!-\!m},
 \end{equation}
we obtain that for even $N$
\begin{subequations}
 \begin{equation} \ket{\Phi(\xi)}=\Big({1+\sqrt{N}\tan\frac{\xi}{2}\,\tilde{c}_0^\dagger}\Big)\cos^2\frac{\xi}{2}\prod_{m=1}^{m<\frac{N}{2}}
\left(\cos^2\frac{\xi}{2}+i\sin^2\frac{\xi}{2}\cot\frac{\pi m}{N}
\tilde{c}^\dagger_m \tilde{c}^\dagger_{N\!-\!m}\right)\ket{\Omega},
 \end{equation}
 whereas
 for odd $N$
 \begin{equation}
\ket{\Phi(\xi)}=\Big({1+\sqrt{N}\tan\frac{\xi}{2}\,\tilde{c}_0^\dagger}\Big)
\cos\frac{\xi}{2}\prod_{m=1}^{m<\frac{N}{2}}
\left(\cos^2\frac{\xi}{2}+i\sin^2\frac{\xi}{2}\cot\frac{\pi m}{N}
\tilde{c}^\dagger_m \tilde{c}^\dagger_{N\!-\!m}\right)\ket{\Omega}.
 \end{equation}
 \end{subequations}
 Therefore, the overlap of $\ket{\Psi_{0}(r,h)}$ with $\ket{\Phi(\xi)}$ for
 even  $N$ is
 \begin{equation} \ipr{\Psi_{0}(r,h)}{\Phi(\xi)}=\sqrt{N}\sin\frac{\xi}{2}\prod_{m=1}^{m<\frac{N}{2}}
\left(\cos\theta_m^{(0)}(r,h)\,\cos^2\frac{\xi}{2}+\sin\theta_m^{(0)}(r,h)\,\sin^2\frac{\xi}{2}\cot\frac{\pi
m}{N}\right),
 \end{equation}
whereas for odd $N$
 \begin{equation} \ipr{\Psi_{0}(r,h)}{\Phi(\xi)}=\sqrt{N}\sin\frac{\xi}{2}\cos\frac{\xi}{2}\prod_{m=1}^{m<\frac{N}{2}}
\left(\cos\theta_m^{(0)}(r,h)\,\cos^2\frac{\xi}{2}+\sin\theta_m^{(0)}(r,h)\,\sin^2\frac{\xi}{2}\cot\frac{\pi
m}{N}\right).
 \end{equation}

Summarizing the above derivations: with $\ket{\Psi_0}$ ($\ket{\Psi_{1/2}}$)
denoting the lowest state in the odd (even) fermion-number sector, we arrive
at the overlaps
\def\myaa{b}
\begin{equation}
\label{eqn:overlap}
\ipr{\Psi_{\myaa}(r,h)}{\Phi(\xi)}
=
f^{(\myaa)}_{N}(\xi)\prod_{m=1-2\myaa}^{m<\frac{N-1}{2}}
\left[
\cos\theta^{(\myaa)}_{m}(r,h)\cos^2({\xi}/{2})+
\sin\theta^{(\myaa)}_{m}(r,h)\sin^2({\xi}/{2})
\cot(k_{m,N}^{(\myaa)}/2)
\right],
\end{equation}
with
\begin{subequations}
\label{eqn:f}
\begin{eqnarray}
&& k_{m,N}^{(\myaa)}
\equiv
\frac{2\pi}{N}(m+{\myaa}),
\ \ \ \tan2\theta^{(\myaa)}_{m}(r,h)
\equiv
r\sin k_{m,N}^{(\myaa)}\big/
(h\!-\!\cos k_{m,N}^{(\myaa)});
\\
&&
f^{(1/2)}_{N}(\xi)
\equiv 1,
\ \ \
f^{(0)}_{N}(\xi)
\equiv
\sqrt{N}\sin({\xi}/{2})\cos({\xi}/{2}),
\
(\mbox{$N$ even});
\\
&&
f^{(1/2)}_{N}(\xi)
\equiv
\cos({\xi}/{2}),
\ \ \
f^{(0)}_{N}(\xi)
\equiv
\sqrt{N}\sin({\xi}/{2}),
\ \ \,\,
(\mbox{$N$  odd});
\end{eqnarray}
\end{subequations}
where $\myaa=0,1/2$ and $m\in[0,N-1]$ is the (integer) momentum index. The
above results are {\it exact\/} for arbitrary $N$, obtained with periodic
boundary conditions on spins rather than in the so-called $c$-cyclic
approximation~\cite{LiebSchultzMattis61}. Given these overlaps, we can readily
obtain the entanglement of the ground state, the first excited state, and any
linear superposition, $\cos\alpha\ket{\Psi_0}+\sin\alpha\ket{\Psi_1}$ of the
two lowest states, for arbitrary $(r,h)$ and $N$, by maximizing the magnitude
of the overlap with respect to the single, real parameter $\xi$.

We remark that with the lowest states explicitly expressed in terms of the
fermionic language, their fidelity measure~\cite{fidelity} can be easily
evaluated. Our focus here is the geometric measure of entanglement, which we
derive in the next sections.
\section{Entanglement in the transverse-field XY chains}
\label{sec:EntMain}
\begin{figure}
\psfrag{h}{$h$}
\psfrag{r}{$r$}
{\psfrag{E}{${\cal E}_{N} \ $}
\centerline{\psfig{figure=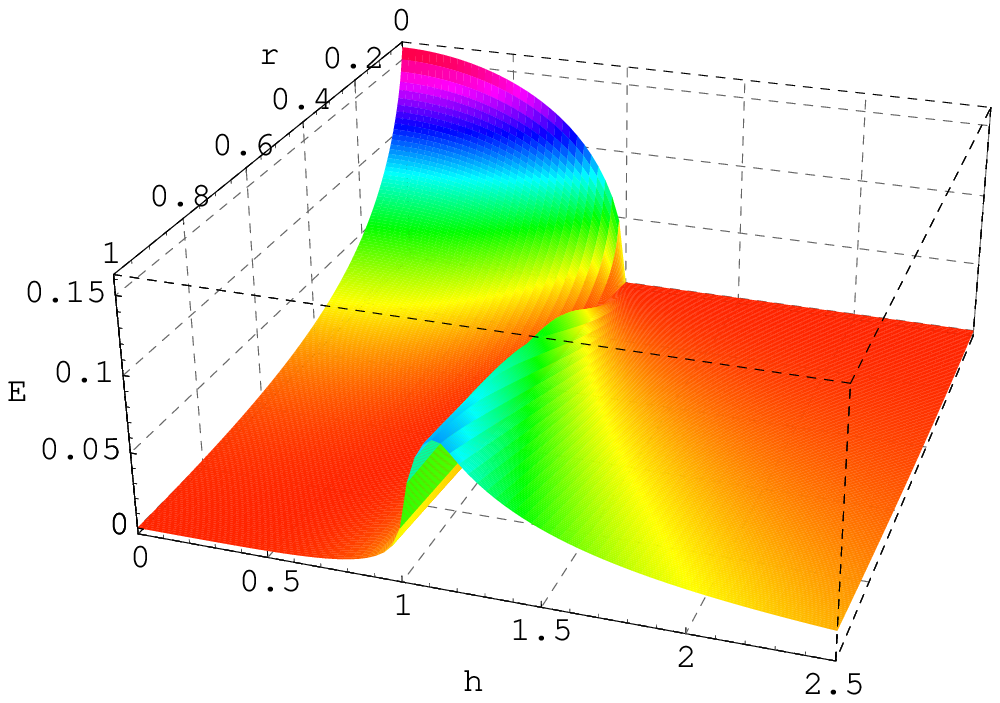,width=12cm,angle=0}}}
\vspace{0.5cm}
  \centerline{\psfig{figure=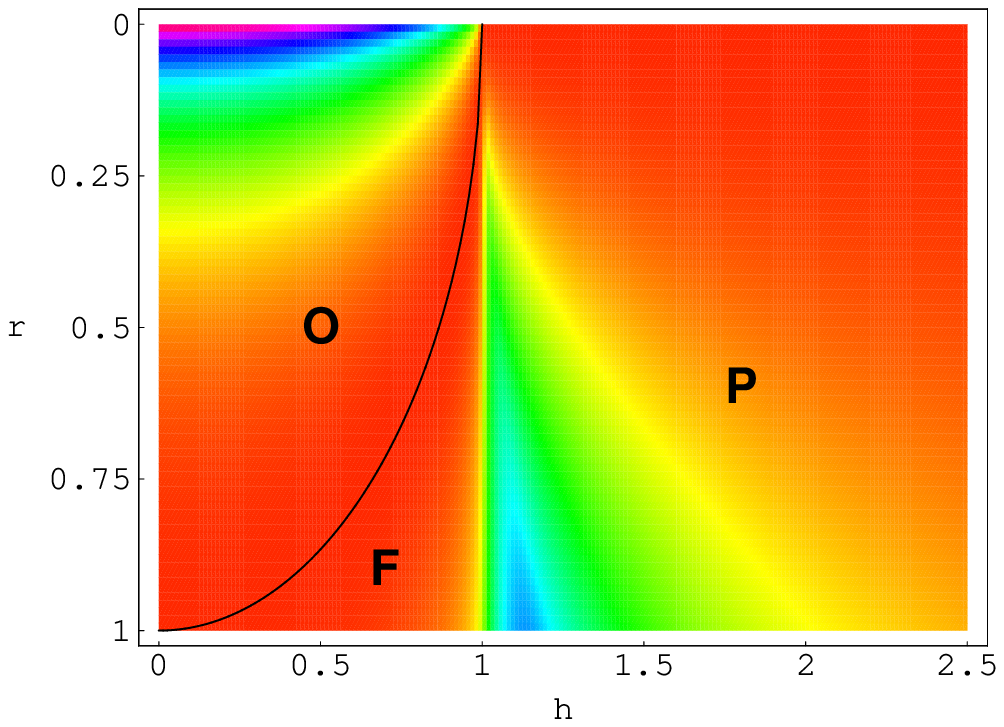,width=12cm,angle=0}}
\caption{(Color online) Entanglement density (upper) and phase
diagram (lower) vs.~$(r,h)$ for the XY model with $N=10^4$ spins,
which is essentially in the thermodynamic limit. There are three
phases: {\bf O}: ordered oscillatory, for $r^2+h^2<1$ and $r\ne 0$;
{\bf F}: ordered ferromagnetic, between $r^2+h^2>1$ and $h<1$; {\bf
P}: paramagnetic, for $h>1$. As is apparent, there is a sharp rise
in the entanglement across the line $h=1$, which signifies a quantum
phase transition. The arc $h^2+r^2=1$, along which the entanglement
density is zero (see also Fig.~\ref{fig:Ent1000}), separates phases
{\bf O} and {\bf F}. Along $r=0$ lies the XX model, which belongs to
a different universality class from the anisotropic XY model. }
\label{fig:XYEnt10000}
\end{figure}

\begin{figure}
\psfrag{h}{$h$}\psfrag{E}{${\cal E}(h)$}\psfrag{D}{${\cal E}'(h)$}
\centerline{\psfig{figure=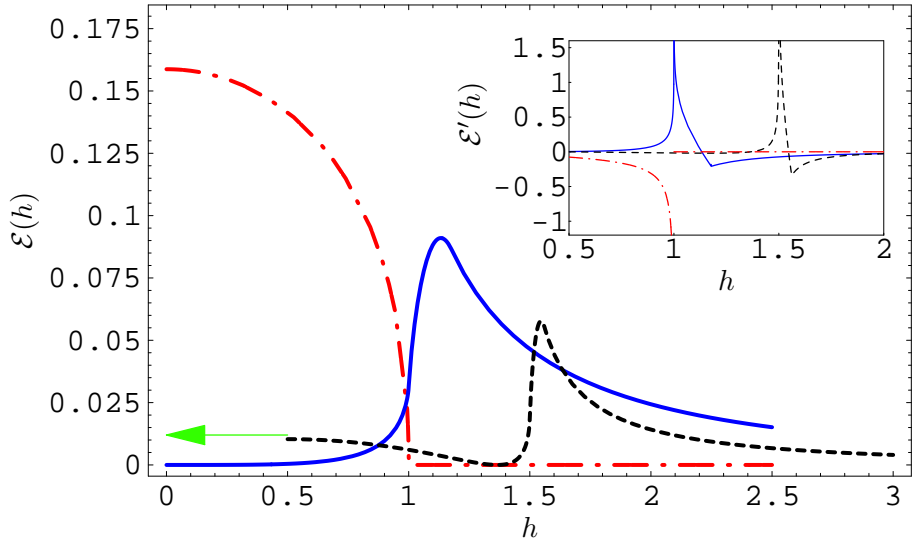,width=12cm,angle=0}}
\caption{(Color online) Entanglement density and its $h$-derivative
(inset) for the ground state of three systems at $N=\infty$. Solid
line: Ising ($r=1$) limit; dashed line: anisotropic ($r=1/2$) XY
model; dash-dotted line: ($r=0$) XX model. For the sake of clarity,
the XY-case curves are shifted to the right by 0.5, indicated by the
arrow. For the $r=1/2$ case, at  the entanglement density vanishes
at $h=\sqrt{1-r^2}$, which is a general property for the anisotropic
XY model. Note that whilst the entanglement itself has a nonsingular
maximum at $h\approx 1.1$ (Ising), $h\approx 1.04$ ($r=1/2$ XY), and
$h=0$ (XX), respectively, it has a singularity at the quantum
critical point at $h=1$, as revealed by the divergence of its
derivative.} \label{fig:Ent1000}
\end{figure}

 The formulas regarding the overlaps of $\ket{\Psi_b}$'s
with product state $\ket{\Phi(\xi)}$ [in Eqs.~(\ref{eqn:overlap})
and~(\ref{eqn:f})] contain all the results that we shall explore shortly,
including both finite-size and infinite-size limits. By analyzing the
structure of Eq.~(\ref{eqn:overlap}), we find that the global entanglement
does provide information on the phase structure and critical properties of the
quantum spin chains.  Two of features, as captured in
Figs.~\ref{fig:XYEnt10000} and~\ref{fig:Ent1000}, are: (i)~although the
entanglement itself is, generically, not maximized at the quantum critical
line in the $(r,h)$ plane, {\it the field-derivative of the entanglement
diverges as the critical line $h=1$ is approached\/}; and (ii)~the
entanglement {\it vanishes\/} at the disorder line $r^2+h^2=1$, which
separates the oscillatory and ferromagnetic phases.

As is to be expected, at finite $N$ the two lowest states $\ket{\Psi_0}$ (with
energy $E^{(0)}_0$) and $\ket{\Psi_{1/2}}$ (with energy $E^{(1/2)}_0$)
featuring in Eq.~(\ref{eqn:overlap}) do not spontaneously break the ${\rm
Z}_2$ symmetry. However, in the thermodynamic limit they are degenerate for
$h\le 1$, and linear combinations are also ground states. The question then
arises as to whether linear combinations that explicitly break ${\rm Z}_2$
symmetry, i.e., the physically relevant states with finite spontaneous
magnetization, show the same entanglement properties. In fact, we see from
Eq.~(\ref{eqn:overlap}) that, in the thermodynamic limit, overlaps for
$\ket{\Psi_0}$ and $\ket{\Psi_{1/2}}$ are identical, up to the prefactors
$f^{(0)}_N$ and $f^{(1/2)}_N$. These prefactors do not contribute to the
entanglement density, and the entanglement density is therefore the same for
both $\ket{\Psi_0}$ and $\ket{\Psi_{1/2}}$. It  further follows that, in the
thermodynamic limit, the results for the entanglement density are insensitive
to the replacement of a symmetric ground state by a broken-symmetry one.

One the other hand, entanglement of finite-size systems does depend on the
exact ground state(s) and which superposition to take when there is
degeneracy. For instance, the ground state switches between $\ket{\Psi_0}$ and
$\ket{\Psi_{1/2}}$ (see Fig.~\ref{fig:gxDE}), and the ground-state
entanglement for a finite system can be broken into pieces (see
Fig.~\ref{fig:gxEntComp}). Furthermore, we shall also investigate the
entanglement of the superposition:
$\cos\theta\ket{\Psi_{1/2}}+\sin\theta\ket{\Psi_{0}}$.

\subsection{Entanglement for finite spins}
\label{subsec:finite}
\begin{figure}
\psfrag{E}{${E}_{log_2}$}
\centerline{\psfig{figure=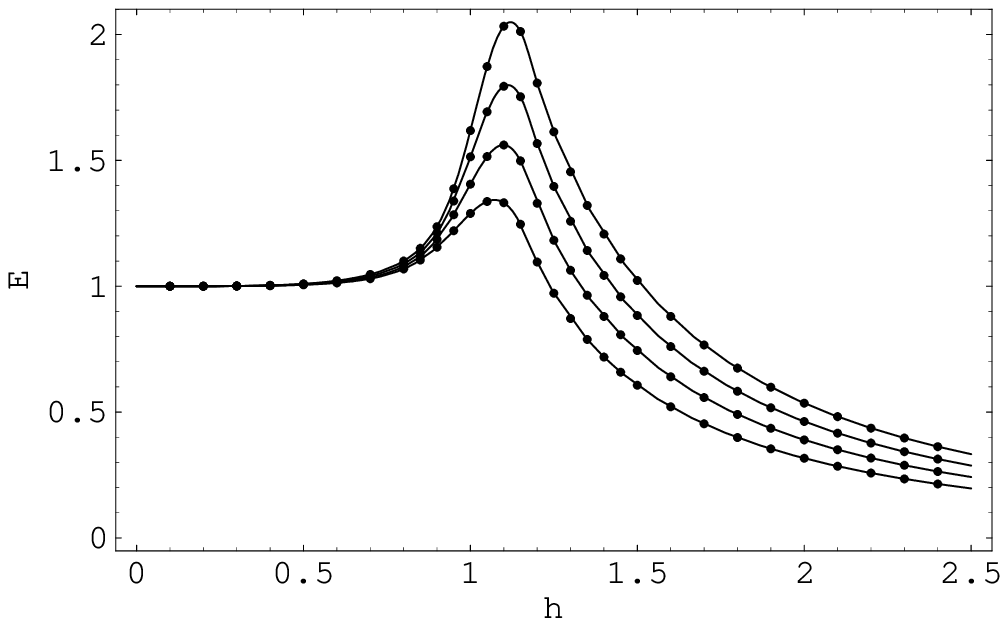,width=10cm,angle=0}}
\psfrag{E}{${E}_{log_2}$}
\centerline{\psfig{figure=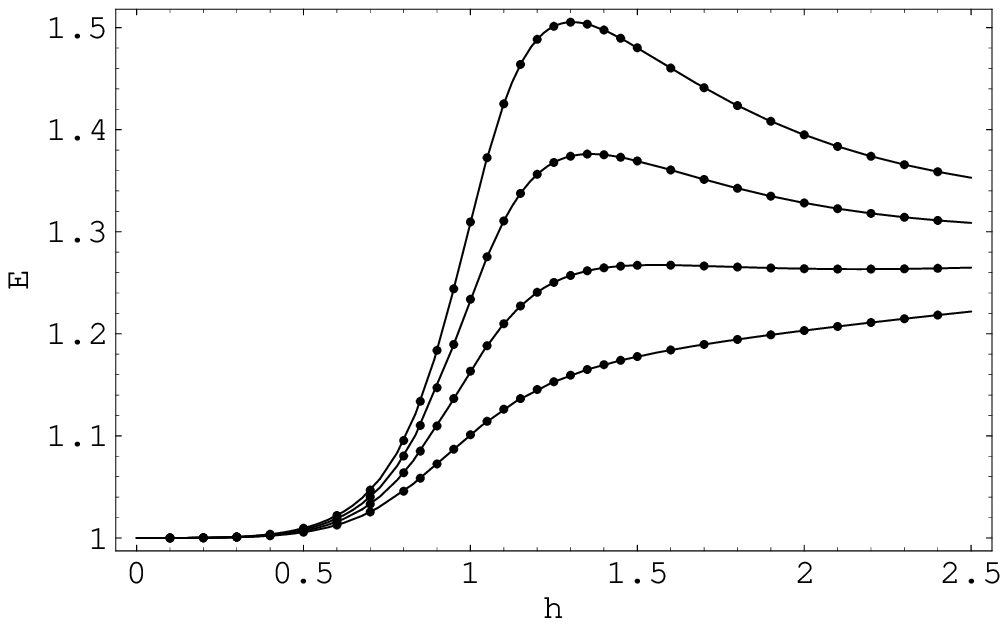,width=10cm,angle=0}}
\caption{Total entanglement vs.~magnetic field $h$ for ground (upper panel) and first excited (lower panel) states with small $N=13,16,19,22$ (from bottom to top) for the transverse Ising model ($r=1$). The numerical results are shown as discrete points whereas
the analytical results are shown as lines. This demonstrates that the
analytical results are exact, even for small $N$, both even and odd. }
\label{fig:4plot}
\end{figure}

Before we discuss the thermodynamic limit of the
entanglement density, we compare the entanglement obtained via the results in
Eq.~(\ref{eqn:overlap}) and that obtained via numerically diagonalizing the
Hamiltonian and calculating the maximal overlap. In Figure~\ref{fig:4plot} the
results via each method are shown for the Ising case ($r=1$) with small
numbers of spins ($N=13$ through $22$), it is seen that our analytical results
are exact even for small $N$, both even and odd.

\begin{figure}
\psfrag{E}{${E}_{log_2}$}
\centerline{\psfig{figure=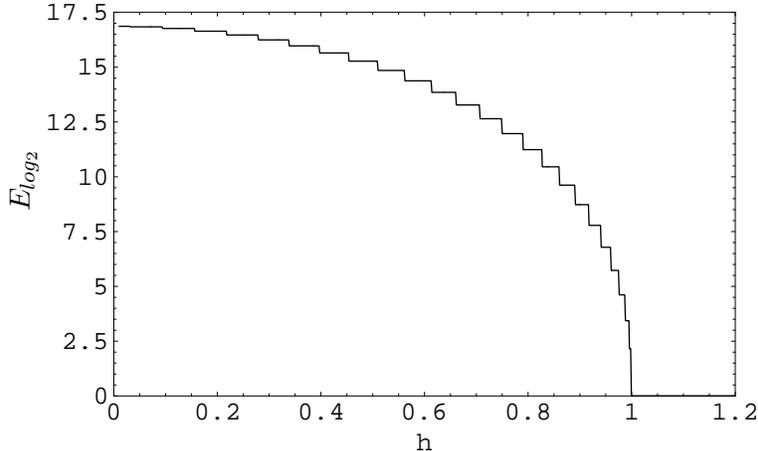,width=10cm,angle=0}} \caption{Total
entanglement vs.~magnetic field $h$ for the ground state of XX chain with
$N=100$ qubits.   } \label{fig:XXN100}
\end{figure}

As another example, we show in Fig.~\ref{fig:XXN100} the total entanglement for
the ground state of the XX chain with $N=100$ spins. The entanglement varies
with $h$ stepwise. This makes sense as at $h=0$, the ground state has continuous
symmetry in XY plane, and hence has high entanglement. As $h$ increases, the
z-component total angular momentum increases by one once the increase exceeds
certain threshold, breaking the continuous
symmetry, and hence the entanglement decreases stepwise until at $h=1$ when all
the spins are pointing in the z-direction. In the limit $N\rightarrow\infty$ the
change in the step per spin becomes infinitesimal and the curve becomes smooth
shown in Fig.~\ref{fig:Ent1000} as expected.

\begin{figure}
\psfrag{E}{${E}_{log_2}$}
\centerline{\psfig{figure=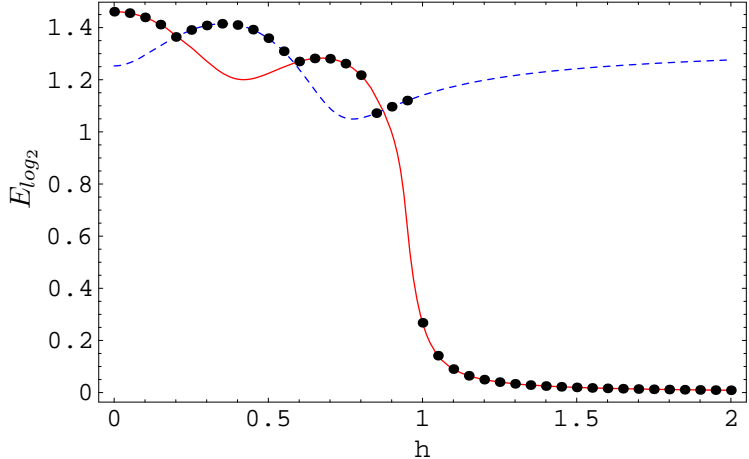,width=10cm,angle=0}} \caption{(Color
online) Total entanglement vs.~magnetic field $h$ of the XY chain with $r=0.2$
and $N=8$ for the ground state (black dots), the minimum energy state in
odd-fermion sector $E_0^{(0)}$ (dashed blue curve) and even-fermion sector
$E_0^{(1/2)}$ (sold red). We see that the ground state, and hence its
entanglement, switches between these two states; see also Fig.~\ref{fig:gxDE}.
} \label{fig:gxEntComp}
\end{figure}

As a final example in this section, we illustrate the strange
feature of piecewise continuity for finite-system ground-state
entanglement. Take $r=0.2$ and $N=8$ for example. The ground state
switch between the two states labeled by the energy $E^{(0)}_0$ (odd
fermion number) and $E^{(1/2)}_0$ (even fermion number), as already
shown in Fig.~\ref{fig:gxDE}. In Fig.~\ref{fig:gxEntComp}, we show
the entanglement for these two states (solid red and dashed blue
dashed curves), as well as the ground-state entanglement obtained
from exact diagonalization (shown as dots). The ground state for
this system has a piecewise continuity in its entanglement. A
similar behavior has also been observed in the context of the
hierarchical geometric
entanglement~\cite{BlasoneDellAnnoDeSienaIlluminati09,BlasoneDellAnnoDeSienaIlluminati10}.
We remark that when one considers the entanglement per site, the
difference between the entanglement density of the two states
vanishes in the thermodynamic limit, as we shall see below.
\subsection{Entanglement density of infinite chains}
\label{subsec:infinite}
 From Eq.~(\ref{eqn:overlap}) it follows that the
thermodynamic limit of the entanglement density is given by
\begin{eqnarray}
\label{eqn:Erh}
{\cal E}(r,h)
=
-\frac{2}{\ln2}\max_\xi
\int_0^{\frac{1}{2}}d\mu\,
\ln\left[\cos\theta(\mu,r,h)\cos^2({\xi}/{2})
+\sin\theta(\mu,r,h)\sin^2({\xi}/{2})\cot\pi\mu\right],
\label{eq:infNent}
\end{eqnarray}
where $\tan 2\theta(\mu,r,h)\equiv r\sin 2\pi\mu /(h-\cos 2\pi\mu)$.

Figure~\ref{fig:Ent1000} shows the thermodynamic limit of the
entanglement density ${\cal E}(r,h)$ and its $h$-derivative in
the ground state, as a function of $h$ for three values of $r$,
i.e., three slices through the surface shown in
Fig.~\ref{fig:XYEnt10000}.
As the $r=1$ slice shows, in the Ising
limit the entanglement density is small for both small and large $h$.
It increases with $h$ from zero, monotonically, albeit very slowly for
small $h$, then swiftly rising to a maximum at $h\approx 1.13$
before decreasing monotonically upon further increase of $h$,
asymptotically to zero.  The entanglement maximum {\it does not\/} occur
at the quantum critical point.  However, the derivative of
the entanglement with respect to $h$ {\it does\/} diverge at the critical
point $h=1$, as shown in the inset.
The slice at $r=1/2$ (for clarity, shifted
 half a unit to the right) shows qualitatively similar
behavior, except that it is finite (although small) at $h=0$, and starts out
by decreasing to a shallow minimum of zero at $h=\sqrt{1-r^{2}}$. By contrast,
the slice at $r=0$ (XX) starts out at $h=0$ at a maximum value of  $1- 2
\gamma_C/(\pi \ln 2)\approx 0.159$. (where $\gamma_C\approx 0.9160$ is the
{\it Catalan\/} constant), the globally maximal value of the entanglement over
the entire $(r,h)$ plane.
For larger $h$ it falls monotonically until it vanishes at $h=1$,
remaining zero for larger $h$.

 Let us analyze the behavior of the entanglement derive for $r\ne0$ (Ising universality class) case first.
\subsubsection{Divergence of entanglement-derivative for the anisotropic XY
models} \label{subsub:DivXY} The starting point is Eq.~(\ref{eqn:Erh}), in
which there is a maximization over the variable $\xi$. The function to be
maximized is
\begin{equation}
F(\xi,r,h)\equiv\int_0^{\frac{1}{2}}d\mu\,
\ln\left[\cos\theta(\mu,r,h)\cos^2({\xi}/{2})
+\sin\theta(\mu,r,h)\sin^2({\xi}/{2})\cot\pi\mu\right].
\end{equation}

To find the stationarity condition, we demand the derivative with
respect to $\xi$ vanishes ($\partial_\xi
F(\xi,r,h)\Big|_{\xi=\xi*}=0$), where
\begin{equation}
\label{eqn:stationarity}
\partial_\xi F(\xi,r,h)\Big|_{\xi=\xi*}=-\frac{1}{2}\sin\xi \int_0^{\frac{1}{2}}d\mu\,
\frac{\cos\theta(\mu,r,h)-\sin\theta(\mu,r,h)\cot\pi\mu}{
\cos\theta(\mu,r,h)\cos^2({\xi}/{2})
+\sin\theta(\mu,r,h)\sin^2({\xi}/{2})\cot\pi\mu}\Big|_{\xi=\xi*}.
\end{equation}
Denote by $\xi^*(h)$ the solution for fixed $r$. Then the  field-derivative of
the entanglement is
\begin{equation}
\partial_h {\cal E}(r,h)= \frac{-2}{\ln2}\partial_h F(\xi^*(h),h)
= \frac{-2}{\ln2}\left[ \frac{\partial \xi^*(h)}{\partial h}
\partial_{\xi}F(\xi,h)\Big|_{\xi^*}+\partial_h F(\xi^*,h)\right]=
\frac{-2}{\ln2}\partial_h F(\xi^*,h),
\end{equation}
where the first term in the square bracket vanishes identically due to the
condition~(\ref{eqn:stationarity}). Thus (dropping the * on $\xi$ for
convenience),
\begin{eqnarray}
\!\!\!\!\!\!\!\!\!\!\!\!\partial_h {\cal E}(r,h)&=&- \frac{2}{\ln2}\partial_h F(\xi^*,h)\\
\!\!\!\!\!\!\!\!\!\!\!\!\label{eqn:EB}
&=&-\frac{2}{\ln2}\int_0^{\frac{1}{2}}d\mu\,
\frac{\partial_h\cos\theta(\mu,r,h)\cos^2({\xi}/{2})
+\partial_h\sin\theta(\mu,r,h)\sin^2({\xi}/{2})\cot\pi\mu}{
\cos\theta(\mu,r,h)\cos^2({\xi}/{2})
+\sin\theta(\mu,r,h)\sin^2({\xi}/{2})\cot\pi\mu}.
\end{eqnarray}
Recall that $\tan 2\theta(\mu,r,h)\equiv r\sin 2\pi\mu /(h-\cos 2\pi\mu)$ and
thus
\begin{subequations}
\begin{eqnarray}
&&\cos\theta=\sqrt{(1+\cos2\theta)/2}, \  \sin\theta=\sqrt{(1-\cos2\theta)/2}, \\
&&\cos2\theta(\mu,r,h)= \frac{h-\cos 2\pi\mu}{\sqrt{(r\sin 2\pi\mu)^2 +(h-\cos
2\pi\mu)^2}}.
\end{eqnarray}
\end{subequations}
Putting everything in Eq.~(\ref{eqn:EB}), we get
\begin{eqnarray}
\partial_h {\cal E}(r,h)&=&-\frac{r}{\ln2}\int_0^{\frac{1}{2}}d\mu\,
\frac{\sin 2\pi\mu}{(r\sin 2\pi\mu)^2 +(h-\cos 2\pi\mu)^2} \nonumber\\
&&\!\!\!\!\!\!\!\!\!\!\!\!\frac{\sqrt{\sqrt{\phantom{AAA}}-(h-\cos2\pi\mu)}\cos^2(\xi/2)-\sqrt{\sqrt{\phantom{AAA}}+(h-\cos2\pi\mu)}\sin^2(\xi/2)\cot\pi\mu}{\sqrt{\sqrt{\phantom{AAA}}+(h-\cos2\pi\mu)}\cos^2(\xi/2)+\sqrt{\sqrt{\phantom{AAA}}-(h-\cos2\pi\mu)}\sin^2(\xi/2)\cot\pi\mu},
\end{eqnarray}
where $\sqrt{\phantom{AAA}}\equiv\sqrt{(r\sin 2\pi\mu)^2 +(h-\cos
2\pi\mu)^2}$.

We aim to explore the behavior near $h=1$. First consider $h>1$ and define
$\epsilon\equiv h-1$, which is the deviation from the critical point. Make the
change of variables $t=h-\cos2\pi\nu$, giving lower and upper limits
$\epsilon$ and $2+\epsilon$, respectively. We further shift the integration
variable by $\epsilon$, arriving at
\begin{eqnarray}
\label{eqn:EB2}
\partial_h {\cal E}(r,h)&=&-\frac{r}{2\pi\ln2}\int_0^{{2}}d\,t
\frac{1}{(1-r^2)t^2+ 2(r^2+\epsilon)t+\epsilon^2}  \nonumber\\
&&\!\!\!\!\!\!\!\!\!\!\!\!\frac{\sqrt{t}\sqrt{\sqrt{\phantom{AAA}}-(t+\epsilon)}\cos^2(\xi/2)-\sqrt{\sqrt{\phantom{AAA}}+(t+\epsilon)}\sin^2(\xi/2)\sqrt{2-t}}{\sqrt{t}\sqrt{\sqrt{\phantom{AAA}}+(t+\epsilon)}\cos^2(\xi/2)+\sqrt{\sqrt{\phantom{AAA}}-(t+\epsilon)}\sin^2(\xi/2)\sqrt{2-t}},
\end{eqnarray}
where $\sqrt{\phantom{AAA}}=\sqrt{(1-r^2)t^2+ 2(r^2+\epsilon)t+\epsilon^2}$.

As we inspect the limit $h\rightarrow1$ or $\epsilon\rightarrow0$, we see that
the above expression diverges, with the contribution coming from $t$ small,
i.e., infrared divergence. Large $t$($\le2$) does not contribute to the
divergence. Note further that only the second term in the numerator
contributes to the divergence. We then proceed to evaluate the integral by
separating it into two parts:
\begin{equation}
\int_0^2=\int_0^\delta +\int_\delta^2,
\end{equation}
with $\delta\ll 1$. In the first region we only need to keep $t$ to first
order at most. Also noting that for $t,\epsilon \ll 1$, the first term in the
denominator is much smaller than the second term, we get
\begin{eqnarray}
-\frac{r}{2\pi\ln2}\int_0^{{\delta}}d\,t
\frac{-1}{2r^2(t+\frac{\epsilon^2}{2r^2})}
\frac{\sqrt{\sqrt{2(r^2+\epsilon)t+\epsilon^2}+(t+\epsilon})}{\sqrt{\sqrt{2(r^2+\epsilon)t+\epsilon^2}-(t+\epsilon})}.
\end{eqnarray}
Next, we simplify the second term (ignoring $\epsilon$ when there is no danger
in doing so)
\begin{equation}
\frac{\sqrt{\sqrt{2(r^2+\epsilon)t+\epsilon^2}+(t+\epsilon})}{\sqrt{\sqrt{2(r^2+\epsilon)t+\epsilon^2}-(t+\epsilon})}=\frac{\sqrt{t+\frac{\epsilon^2}{2r^2}}}{\sqrt{t}}+\frac{t+\epsilon}{\sqrt{2r^2
t}}.
\end{equation}
Observing that only the first term on the right-hand side contributes to the
divergence, we have that the divergent part is
\begin{eqnarray}
-\frac{r}{2\pi\ln2}\int_0^{{\delta}}d\,t
\frac{-1}{2r^2(t+\frac{\epsilon^2}{2r^2})}
\frac{\sqrt{t+\frac{\epsilon^2}{2r^2}}}{\sqrt{t}}=
\frac{1}{4r\pi\ln2}\int_0^{{\delta}2 r^2/\epsilon^2}d\,t \frac{1}{\sqrt{t+1}}
\frac{1}{\sqrt{t}}.
\end{eqnarray}
The divergence part is then (for ${\delta}2 r^2/\epsilon^2\gg 1$)
\begin{equation}
\frac{1}{4r\pi\ln2}2\,{\sinh}^{-1}\sqrt{{\delta}2
r^2/\epsilon^2}\approx\frac{1}{2r\pi\ln2} \ln \left(2\sqrt{{\delta}2
r^2/\epsilon^2}\right)=-\frac{1}{2r\pi\ln2} \ln \epsilon+\frac{\ln
\big(2\sqrt{\delta 2 r^2}\big)}{2r\pi \ln2}.
\end{equation}
As the integral~(\ref{eqn:EB2}) does not depend on the choice of $\delta$, the
part that involves $\delta$ must be cancelled by the second half of the
integration $\int_\delta^2$, which can be verified by direct evaluation.
Therefore, for $h$ very close to $h_c=1$, we have
\begin{equation}
\frac{\partial{\cal E}}{\partial h} \approx -\frac{1}{2\pi r \ln2}\ln|h-1|.
\end{equation}

In deriving the above divergence form, we have assumed that $r\ne 0$. Similar
consideration can be applied to the case when $h$ approaches 1 from below, and
the behavior is the same.

\subsubsection{Divergence of entanglement-derivative for the  XX limit of the
model} \label{subsub:DivXX} We now analyze the $r=0$ isotropic (XX
universality class) case. It turns out that the analysis for this case is much
simpler. To see this, we first note when $r=0$ we have the simplification
\begin{equation}
\cos2\theta(\mu,h)={\rm sgn}(h-\cos2\pi\mu).
\end{equation}
The above expression changes sign when $h=\cos2\pi\nu$. So let us introduce
the variable $\mu_0\equiv (\cos^{-1}h)/(2\pi)$. The expression for the
entanglement density Eq.~(\ref{eqn:Erh}) becomes
\begin{subequations}
\begin{eqnarray}
\!\!\!\!\!\!\!\!\!\!{\cal E}(h) \label{eqn:EB3}
\!\!\!\!\!\!\!\!\!\!&=&-\frac{2}{\ln2}\max_\xi
\int_0^{\frac{1}{2}}d\mu\,
\ln\left[\cos\theta(\mu,r,h)\cos^2({\xi}/{2})
+\sin\theta(\mu,r,h)\sin^2({\xi}/{2})\cot\pi\mu\right]\\
\!\!\!\!\!\!\!\!\!\!&=&-\frac{2}{\ln2}\max_\xi\left[
\int_0^{\mu_0}d\mu\, \ln\sin^2({\xi}/{2})\cot\pi\mu +
\int_{\mu_0}^{\frac{1}{2}}d\mu\, \ln\cos^2({\xi}/{2})\right].
\end{eqnarray}
\end{subequations}
Demanding stationarity with respect to $\xi$ gives the condition
\begin{equation}
[\mu_0- \frac{1}{4}(1-\cos\xi)]\sin\xi=0.
\end{equation}
The solution $\xi=0$ gives the entanglement density for $h\ge 1$. It is
straightforward to see from Eq.~(\ref{eqn:EB3}) that the entanglement density
is identically zero. The solution $\mu_0- \frac{1}{4}(1-\cos\xi)=0$ gives
\begin{equation}
\cos\xi=1-\frac{2}{\pi}\cos^{-1}h.
\end{equation}
This in turn gives the entanglement density for $0\le h\le 1$:
\begin{equation}
\label{eqn:Eh2} {\cal E}(h)
=-\frac{2}{\ln2}\left[\mu_0(h)\ln\frac{2\mu_0(h)}{1-2\mu_0(h)}+\frac{1}{2}\ln(1-2\mu_0(h))+\int_0^{\mu_0(h)}
d\mu\ln \cot\pi\mu\,\right].
\end{equation}
Thus we have derived the entanglement density as a function of the magnetic
field $h$ in the XX limit. The result is shown in Fig.~(\ref{fig:Ent1000}).

 We see from the figure that the entanglement density at $h=0$ is
the highest, which being ${\cal E}(0)=1- 2 \gamma_C/(\pi \ln 2)\approx 0.159$
by evaluating Eq.~(\ref{eqn:Eh2}) at $h=0$. The constant $\gamma_C\approx
0.9160$ is the {\it Catalan\/} constant. The entanglement density decreases
monotonically as $h$ increases until $h=1$ beyond which it becomes zero
identically. This qualitative behavior can be understood as follows. As the
total $z$-component spin is conserved, increasing $h$ simply increases the
z-component spin of the ground state until $h=1$ where all the spins are
aligned with the field, hence there is no entanglement beyond this value of
$h$.

By directly taking the derivative with respect to $h$, we get
\begin{equation}
\partial_h {\cal E}(h)=\frac{1}{\pi\ln2 \sqrt{1-h^2}} \ln\left[\frac{\cos^{-1}h}{\pi-\cos^{-1}h}\sqrt{\frac{1+h}{1-h}}\,\right].
\end{equation}
Near $h\approx 1$, we have (putting $1+h=2$ and evaluating the limit in the
argument of log function)
\begin{equation}
\label{eqn:DivXX}
 \frac{\partial}{\partial h}{\cal E}(0,h)\approx
-\frac{\log_2(\pi/2)}{\sqrt{2}\,\pi} \frac{1}{\sqrt{1-h}},
\qquad (h\to 1^{-}).
\end{equation}
This completes our derivation of the singular behavior of the entanglement
density near the critical points.

\begin{figure}
\psfrag{h}{$h$}
 \psfrag{E}{$\partial{\cal E}_{N}/\partial h
\vert_{h_{{\rm max},N}}$}\psfrag{N}{$\ln N$}
\centerline{\psfig{figure=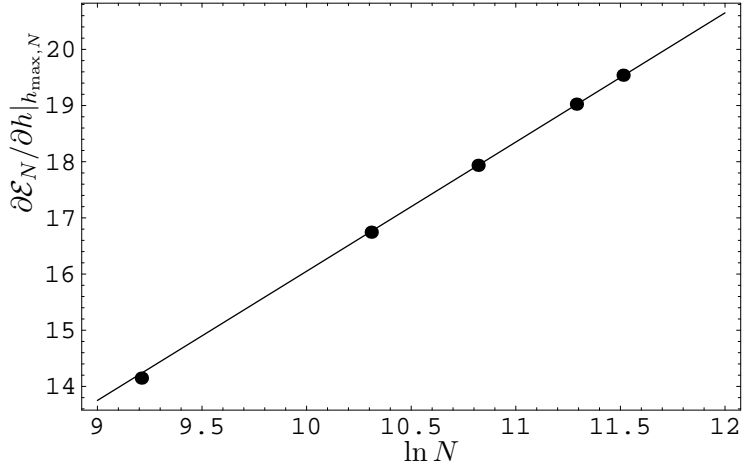,width=10cm,angle=0}}
\caption{Finite-size scaling. $\partial{\cal E}_{N}/\partial h
\vert_{h_{{\rm max},N}}$ vs. $\ln N$, for $N=10^4, 3\times10^4,
5\times10^4, 8\times10^4,$ and  $10^5$ (points) with $r=0.1$. The
solid line represents the fit $\partial{\cal E}_{N}/\partial h
\vert_{h_{{\rm max},N}}\approx 2.30 \ln N -6.95$.}
\label{fig:dEvsNr01}
\end{figure}

\subsubsection{Singularity of entanglement and quantum criticality}
\label{subsub:logscaling} Let us summarize the above derivations of the
singular behavior in the entanglement density and discuss the connection to
quantum criticality, in particular the correlation length critical exponent.

(1) The singular behavior of the field-derivative of the entanglement
density~(\ref{eq:infNent}) in the vicinity of the quantum critical line acts
as (for $r\ne 0$)
\begin{equation}
\label{eqn:EntDiv}
\frac{\partial{\cal E}}{\partial h}
\approx
-\frac{1}{2\pi r \ln2}\ln|h-1|,
\,\,\,
\mbox{for $\vert{h-1}\vert\ll 1$}.
\end{equation}
From the arbitrary-$N$ results~(\ref{eqn:overlap}) of the entanglement we analyze the approach
to the thermodynamic limit, in order
to develop further connections with quantum criticality.
We focus on the exponent $\nu$, which governs the divergence at
criticality of the correlation length: $L_c\sim |h-1|^{-\nu}$.
To do this, we compare the divergence of the slope
$\partial{\cal E}_{N}/\partial h$
(i)~near $h=1$ (at $N=\infty$), given above, and
(ii)~for large $N$ at the value of $h$
for which the slope is maximal (viz.\ $h_{{\rm max},N}$), i.e.,
$\partial{\cal E}_{N}/\partial h
\vert_{h_{{\rm max},N}}\approx
0.230r^{-1}\ln N + {\rm const.}$,
 obtained by analyzing Eq.~(\ref{eqn:overlap})
for various values of $r$; see Fig.~\ref{fig:dEvsNr01} for the example of $r=0.1$ case.
Then, noting that $(2\pi\ln2)^{-1}\approx 0.2296$
and that the logarithmic scaling hypothesis~\cite{Barber83} (or see Appendix~\ref{app:finitesize})
identifies $\nu$ with the ratio of the amplitudes of these divergences,
$0.2296/0.230\approx 1$, we recover the known result that $\nu=1$.
Moreover, from Eq.~(\ref{eqn:znu}) we can extract the value of the
dynamical exponent: $z=1$.

(2) Compared with $r\ne 0$ case, the nature of the divergence of
$\partial{\cal E}/\partial h$ at $r=0$  belongs to a different (XX)
universality class:
\begin{equation}
\label{eqn:EntDivXX}
\frac{\partial}{\partial h}{\cal E}(0,h)\approx
-\frac{\log_2(\pi/2)}{\sqrt{2}\,\pi}
\frac{1}{\sqrt{1-h}},
\qquad (h\to 1^{-}).
\end{equation}
From this divergence, the scaling hypothesis, and the assumption that
the entanglement density is intensive,
we can infer the known result~\cite{Sachdev} that the critical
exponent $\nu=1/2$ for the XX model. Moreover, from Eq.~(\ref{eqn:znu}) we can extract the value of the dynamical exponent $z=2$ for the XX model.

In keeping with the critical features of the XY-model phase diagram,
for any small but nonzero value of the anisotropy,
the critical divergence of the entanglement derivative
is governed by Ising-type behavior.
It is only at the $r=0$ point that the critical behavior of
the entanglement is governed by the XX universality class.
For small $r$, XX behavior ultimately crosses over to Ising behavior.

\subsection{Vanishing of entanglement density along the disorder line}
\label{subsec:disorder}
 We find that along the line $r^{2}+h^{2}=1$ the
entanglement density vanishes in the thermodynamic limit. In fact, this line
exactly corresponds to the boundary separating the oscillatory and
ferromagnetic phases; the boundary can be characterized by a set of ground
states with total entanglement of order unity, and thus of zero entanglement
density. The entanglement density is also able to track the phase boundary
($h=1$) between the ordered and disordered phases. Associated with the quantum
fluctuations accompanying the transition, the entanglement density shows a
drastic variation across the boundary and the field-derivative diverges all
along $h=1$. The two boundaries separating the three phases coalesce at
$(r,h)=(0,1)$, i.e., the XX critical point. Figures~\ref{fig:XYEnt10000} and
\ref{fig:Ent1000} reveal all these features.

 One
extreme limit is the Ising case, i.e., $r=1$ and $h=0$, where the ground state
is either $\ket{\!\!\rightarrow\rightarrow\dots\rightarrow}$ or
$\ket{\!\!\leftarrow\leftarrow\dots\leftarrow}$, both of these being
unentangled. Any superposition of them is also a valid ground state, but it
has entanglement of order unity. In the thermodynamic limit the entanglement
per spin is identically zero. Is this a general feature along the disorder
line? Before we establish this, recall that the energies of the lowest two
levels are given in Eqs.~(\ref{eqn:E0}) and (\ref{eqn:E1/2}). Evaluating them
at $r^2=1-h^2$, we immediately find that  both are $-N$.

Now let us evaluate the expectation value of the Hamiltonian~(\ref{eqn:HXY})
with respect to a separable state with all spins pointing in the same
direction:
\begin{equation}
\langle{\cal H}\rangle =- N \left(\frac{1+r}{2}\langle\sigma^x\rangle^2
+\frac{1-r}{2}\langle\sigma^y\rangle^2+h\langle\sigma^z\rangle\right).
\end{equation}
Denoting $x\equiv\langle\sigma^x\rangle$, $y\equiv\langle\sigma^y\rangle$,
$z\equiv\langle\sigma^z\rangle$, we find that the above expression
achieves its minimum value $-N$ at $r^2+h^2=1$ when
\begin{equation}
(x,y,z)=\left(\pm\sqrt{\frac{2r}{1+r}},0,\sqrt{\frac{1-r}{1+r}}\right).
\end{equation}
Therefore, the separable state satisfying the above conditions is
the ground state, and there is a degeneracy. Hence, along the
disorder line the total entanglement can be at most of order unity
(due to superposition of the above degenerate states) and hence the
entanglement density vanishes. We note that the theory of factorized
ground states has recently been extensively developed and greatly
understood~\cite{Roscilde04,FactorizedPrevious,Factorized}, and it
turns out that they occur quite often and their occurrence is
related to the existence of quantum critical points nearby in the
system parameter.

\subsection{Entanglement of superposition}
\label{subsec:superposition}
\begin{figure}
\psfrag{h}{$h$}
 \psfrag{E}{${E}_{log_2}$}
\centerline{\psfig{figure=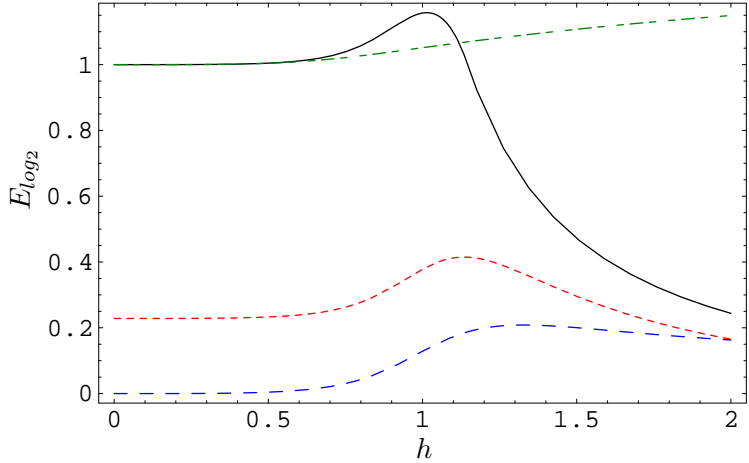,width=10cm,angle=0}}
\caption{(Color online) Total entanglement ($E_{\log_2}$) for an
Ising chain vs. the field $h$ for $N=10$ and various superposition
of $\cos\theta\ket{\Psi_{1/2}}+\sin\theta\ket{\Psi_0}$: $\theta=0$
(solid black), $\theta=\pi/8$ (short-dashed red), $\theta=\pi/4$
(long-dashed blue), and $\theta=\pi/2$ (mixed-dashed green). }
\label{fig:gIsing10}
\end{figure}
In this section, we discuss the entanglement of superposition for the two
lowest states $\ket{\Psi_{1/2}}$ and $\ket{\Psi_0}$:
\begin{equation}
\ket{\Psi(r,h,\theta)}=\cos\theta\ket{\Psi_{1/2}(r,h)}+\sin\theta\ket{\Psi_0(r,h)}.
\end{equation}
Why does this interest us? Take for instance the Ising model, $r=1$. The
states $\ket{\Psi_{1/2}}$ and $\ket{\Psi_0}$, respectively, analytically
continues to (as $h\rightarrow0$)
\begin{subequations}
\begin{eqnarray}
\ket{\Psi_{1/2}(1,0)}&=&(\ket{\!\rightarrow\rightarrow\dots\rightarrow}+\ket{\!\leftarrow\leftarrow\dots\leftarrow})/\sqrt{2}\\
\ket{\Psi_{0}(1,0)}&=&(\ket{\!\rightarrow\rightarrow\dots\rightarrow}-\ket{\!\leftarrow\leftarrow\dots\leftarrow})/\sqrt{2}.
\end{eqnarray}
\end{subequations}
This is a result of unbroken $Z_2$ symmetry and both states possess global
geometric entanglement of unity. In the large system size limit, this symmetry
is usually broken due to fluctuations of a random magnetic field, and the
resulting symmetry-broken states will be unentangled, having zero
entanglement. These symmetry-broken states can be represented as certain
superposition of the original symmetry-unbroken states. But to evaluate the
entanglement for these symmetry-broken states at a different field value $h$,
we need to be able to calculate the entanglement of the superposition, shown
in the above equations.

Fortunately, we have already evaluated the overlap of the two states with the
same product state $\ket{\Phi(\xi)}$, as summarized in
Eqs.~(\ref{eqn:overlap}) and~(\ref{eqn:f}). We only need to compute the sum of
the two overlaps weighted by the coefficients $\cos\theta$ and $\sin\theta$
and maximize the norm of the overlap with respect to the parameter $\xi$. As
an illustration, we have computed the entanglement of superposition for the
Ising model with $N=10$ spins, as shown in Fig.~\ref{fig:gIsing10}. The
resultant entanglement does depend on the detail of the superposition.

The above exercise can be generalized to all values of $r$, $h$, and $N$,
including the infinite limit. If we are interested only in the entanglement
per spin, it is now readily easy to see from Eq.~(\ref{eqn:overlap}) that, in
the thermodynamic limit, overlaps for $\ket{\Psi_0}$ and $\ket{\Psi_{1/2}}$
are identical, up to the prefactors $f^{(0)}_N$ and $f^{(1/2)}_N$. These
prefactors do not contribute to the entanglement density (although they do
matter for the correction in $1/N$), and the entanglement density is therefore
the same for both $\ket{\Psi_0}$ and $\ket{\Psi_{1/2}}$. It  further follows
that, in the thermodynamic limit, the results for the entanglement density are
insensitive to the replacement of a symmetric ground state by a
broken-symmetry one.
\subsection{Finite-size scaling of entanglement at critical points}
\label{subsec:finite-size}
 Recall that we have performed a finite-size scaling
of the entanglement derivative in Sec.~\ref{subsub:logscaling} in order to
extract the critical exponent for the Ising universality class. What about the
finite-size scaling of the entanglement density itself near criticality? This
question was recently raised and addressed by  Shi and
co-workers~\cite{finitesize}, and they have found a scaling of the
entanglement per site ${\cal E}_N$ as
\begin{equation}
\label{eqn:finiteSize} {\cal E}_N \sim {\cal E}_\infty + \frac{b}{N}
+\frac{c}{N^2}.
\end{equation}

With the formulas of entanglement given above, we can address this
question straightforwardly. But as we have discussed earlier, the
correction to the entanglement may depend on which superposition of
degenerate states we use. To avoid the ambiguity, we present the
finite-size scaling at $h=1$ for both states $\ket{\Psi_{1/2}}$
(which is the ground state at $h=1$) and $\ket{\Psi_0}$ at various
values of $r\ne 0$. We shall see that although their ${\cal
E}_\infty$'s are practically the same, the scaling coefficients $b$
and $c$ are different. One interesting observation is that (for
$r\ne 0$) we obtain $b\approx 1.0$ for $\ket{\Psi_{1/2}}$ and
$b\approx0.5$ for $\ket{\Psi_{0}}$. The correction $b/N$ has
recently been shown to related to the Affleck-Ludwig boundary
entropy by St\'ephan and co-workers~\cite{StephanMisguichAlet}.
Furthermore, $c$ is negative for $\ket{\Psi_{1/2}}$ and positive for
$\ket{\Psi_{0}}$. (For $r=0$ XX model at $h=1$, the entanglement
density is identically zero. There is no need to perform the
finite-size scaling.) To get the scaling fit, we use the size $N$
from 100 to 1000, and calculate the respective entanglement for
fitting Eq.~(\ref{eqn:finiteSize}). The results are tabulated in
Table~\ref{tb:1}. Curiously, the $c$'s for $\ket{\Psi_{1/2}}$ are
approximately half and negative of those for $\ket{\Psi_0}$, i.e.,
$c^{(1/2)}\approx - c^{(0)}/2$.
\begin{table}
\begin{tabular}{|l||l|l|l||l|l|l|}
\hline
 &\multicolumn{3}{c|}{$\ket{\Psi_{1/2}}$}&\multicolumn{3}{c|}{$\ket{\Psi_{0}}$}\\
\cline{2-7}
 r&${\cal E}_\infty$&$b$&$c$&${\cal E}_\infty$&$b$&$c$\\
\hline\hline
0.1&0.00426345&1.00266&-10.6643 &0.00425344 &0.512785 & 17.5478\\
0.2&0.00817338&1.00038&-5.39161 &0.00817141 & 0.502356 &10.1401\\
0.3&0.0117754&1.00013&-3.73401  &0.0117747 & 0.500874 &7.23486\\
 0.4&0.0151133&1.00006&-2.91527  & 0.0151129 &0.500444 & 5.71366\\
 0.5&0.0182208&1.00004&-2.42561  & 0.0182206 & 0.500268 &4.78119\\
 0.6&0.0211258&1.00002&-2.09918  & 0.0211256 &0.500181 &4.15147\\
 0.7&0.0238512&1.00002&-1.86569  & 0.0238511 & 0.500131 &3.6975\\
 0.8&0.0264164&1.00001&-1.69019  & 0.0264163 & 0.500101& 3.35452\\
 0.9&0.0288377&1.00001&-1.55332  & 0.0288376 & 0.50008 &3.08608\\
 1.0&0.0311291&1.00001&-1.44349  &0.031129 & 0.500066 &2.87013\\
\hline
\end{tabular}
\caption{Table for the scaling parameters ${\cal E}_\infty$, $b$, and $c$ at
various $r$ but at fixed $h=1$ for $\ket{\Psi_{1/2}}$ and $\ket{\Psi_0}$. The
results are obtained from fitting the number of spins $N$ from 100 to
1000.\label{tb:1}}
\end{table}

\section{Concluding remarks}
\label{sec:conclude} In summary, we have quantified the global entanglement of
the quantum XY spin chain.  This model exhibits a rich phase structure, the
qualitative features of which are reflected by this entanglement measure.
Perhaps the most interesting aspect is the divergence in the field-derivative
of the entanglement as the critical line ($h=1$) is crossed.  The behavior of
the divergence is dictated by the universality class of the model.
Furthermore, in the thermodynamic limit, the entanglement density vanishes on
the disorder line ($r^2+h^2=1$). The structure of the entanglement surface, as
a function of the parameters of the model (the magnetic field $h$ and the
coupling anisotropy $r$), is surprisingly rich. We have also discussed issues
of superposition of entanglement, its independence on the spontaneous breaking
of the $Z_2$ symmetry, and finite-size scaling at critical points

How does ground-state entanglement reflect quantum phase transitions? Can we
have a generic understanding of the connection? Near quantum critical points,
the correlation length generally scales as
\begin{equation}
\xi(h)\sim |h-h_c|^{-\nu},
\end{equation}
with $\nu$ denoting the so-called correlation-length exponent.  For any
density functions, such as free energy density and the entanglement density,
 we expect scaling for the singular part behaves as
(e.g. Ref.~\cite{Goldenfeld})
\begin{equation}
{\cal E}(h)\sim\big(\xi/a\big)^{-d}(c_1 + c_2 \log (\xi/a) + \dots),
\end{equation}
where $d$ is the dimension of the system, $a$ is the smallest length scale,
$c$'s are some constants which may depend on which side of critical point we
are at, and we have included a possible logarithmic correction in its simplest
form. The idea is that near criticality the only length scale is the
correlation length, and any density quantity should scale as 1/volume, i.e.,
$1/(\xi/a)^d$, with $a$ being a unit of length. We then expect that in general
\begin{equation}
\label{eqn:EntScaling} \frac{d {\cal E}(h)}{dh}\Big\vert_{h_c^\pm}\sim \pm
a^d|h-h_c|^{d\,\nu-1}(\tilde{c}_1 + \tilde{c}_2 \log|h-h_c|).
\end{equation}
For the one-dimensional ($d=1$) Ising universality class, $\nu=1$, and there
is a logarithmic divergence, which is consistent with Eq.~(\ref{eqn:EntDiv}).
For the XX universality near $h=1$, there is no logarithmic correction (i.e.,
$c_2=\tilde{c}_2=0$), the correlation length exponent $\nu=1/2$. The resultant
divergence from the above scaling form is consistent with what we have
obtained in Eq.~(\ref{eqn:DivXX}), noting that the coefficient $c_1=0$ for
$h>h_c$.

 We close by pointing towards a deeper connection between the global measure
of entanglement and the correlations among quantum fluctuations. The maximal
overlap~(\ref{eq:lambdamax}) can be decomposed in terms of correlation
functions:
\begin{equation}
\Lambda_{\max}^2
=\frac{1}{2^N}+
\frac{N}{2^N}\max_{|\vec{r}|=1}
\Big\{\langle\vec{r}\cdot\vec{\sigma}_{1}\rangle+\frac{1}{2}
\sum_{j=2}^N\langle\vec{r}\cdot \vec{\sigma}_{1}
\otimes\vec{r}\cdot \vec{\sigma}_{j}\rangle+
\cdots\Big\},\nonumber
\end{equation}
where
translational invariance is assumed and the Cartesian coordinates of $\vec{r}$ can be taken to be
$(\sin\xi,0,\cos\xi)$. The two-point correlations appearing in the
decomposition are related to a bi-partite measure of entanglement, namely, the
concurrence, which shows similar singular behavior~\cite{OsterlohAmicoFalciFazio02} to Eq.~(\ref{eqn:EntDiv}).

In fact, it was elaborated in Ref.~\cite{OrusWei} that a singularity of the
entanglement can come from two types of sources: (i) correlation functions,
${\cal C}^{\alpha,\beta,\gamma,\dots}_{[i,j,k,\dots]}\equiv \langle
{\sigma}_{[i]}^\alpha {\sigma}_{[j]}^\beta{\sigma}_{[k]}^\gamma\dots\rangle$
for the ground state $\ket{\Psi}$, and (ii) parameters $\vec{r}^{\ *[i]}$,
which denote the vectors that maximize the overlap.  For transverse Ising,
which has a standard second-order quantum critical point, (i) correlation
functions ${\cal C}$'s are singular but (ii) optimal parameters $r^*$'s are
not singular and this explains the similar behavior between the GE and the
 concurrence. It can happen that the singularity of the entanglement near a transition can come
 from different types. In Ref.~\cite{OrusWei}, it was realized that
 near the so-called Kosterliz-Thouless point (the isotropic antiferromagnetic point)
 of the XXZ model, the correlation length is finite, but (ii)
$r^*$'s are singular. It is this second point the one that detects
non-analyticity in the wavefunction across the transition.

\medskip \noindent
{\it Acknowledgments\/}---%
The authors acknowledge many valuable discussions with D. Das, S.
Mukhopadyay, and R. Or\'us. This work was supported by NSERC and
MITACS (TCW), by NSF DMR-0644022-CAR (SV) and by the U.S. Department
of Energy,
        Division of Materials Sciences under Award No. DE-FG02-07ER46453, through
        the Frederick Seitz Materials Research Laboratory at the University of Illinois at
        Urbana-Champaign (PMG).

\appendix
\section{Finite-size scaling} \label{app:finitesize} The discussion here
follows Barber~\cite{Barber83}.
In the vicinity of the bulk critical temperature $T_C$ the behavior of a
system should depend on $y\equiv L/\xi(T)$, where $\xi(T)$ is the bulk
correlation length and $L$ is the characteristic length of the system. How
does the divergence of certain thermodynamic quantities emerge as the system
size $L$ grows?

\subsection{Algebraic divergence} Assume that some thermodynamic quantity at
$L\rightarrow\infty$ diverges as $t\equiv(T-T_C)/T_C\rightarrow\infty$:
\begin{equation}
P_\infty(T)\sim C_\infty t^{-\rho}.
\end{equation}
Finite-size scaling hypothesis asserts that for finite $L$ and $T$ near $T_C$,
\begin{equation}
P_L(T)\sim l^\omega Q_P(l^{{1}/{\nu}}\tilde{t}), \ \ l\rightarrow\infty,\
\tilde{t}\rightarrow 0,
\end{equation}
where $l\equiv L/a$ ($a$ is some microscopic length), $\tilde{t}\equiv
[T-T_C(L)]$. The exponent $\omega$ can be determined by the requirement that
the $P_L(T)$ reproduces $P_\infty(T)$ as $l\rightarrow\infty$. Thus,
\begin{equation}
Q_P(x)\sim C_\infty x^{-\rho}, \ \ x\rightarrow\infty,
\end{equation}
and $\omega=\rho/\nu$. We consider the case that the finite system does not
exhibit a true transition, then
\begin{equation}
Q_P(x)\rightarrow Q_0, \ \ x\rightarrow 0.
\end{equation}
From this we have that at the peak or rounding temperature $T_m^*(l)$ (where
$P_L$ reaches the maximum or deviates significantly from the bulk value)
\begin{equation}
P_L(T_m^*(l))\sim Q_0 l^{\rho/\nu}, \ \ l\rightarrow\infty.
\end{equation}
This means that the behavior of a thermodynamic quantity varies with the
system size is determined by the bulk critical exponent.
\subsection{Logarithmic divergence} Now assume the thermodynamic quantity
$P(T)$ diverges logarithmically as
\begin{equation}
P_\infty(T)\sim C_\infty \ln t, \ \ t\rightarrow0,
\end{equation}
as in the field-derivative of the entanglement density for anisotropic XY spin
chains. The finite-size scaling hypothesis in this case is to assume
\begin{equation}
P_L(T)-P_L(T_0)\sim Q_P(l^{1/\nu}\tilde{t})-Q_P(l^{1/\nu}\tilde{t}_0),
\end{equation}
where $T_0$ is some non-critical temperature and $\tilde{t}_0\equiv
(T_0-T_C(L))/T_C$. The hypothesis has to recover the $l\rightarrow\infty$
limit at fixed $T$, which requires
\begin{equation}
Q_P(x)\sim C_\infty\ln x, \ \ x\rightarrow\infty.
\end{equation}
Thus in the limit $\tilde{t}\rightarrow0$ at fixed large $l$, we have
\begin{equation}
P_L(T_C(L))\sim -\frac{C_\infty}{\nu}\ln l + O(1), 
\end{equation}
if $Q_P(x)=O(1)$ as $x\rightarrow 0$. This allows us to obtain the exponent
$\nu$ by analyzing how the divergence develops as the system size $l$ (which
is $N$ in our spin-chain entanglement problem) increases.


\begin{thebibliography}{99}
\bibitem{footnote:ansatz}
We have confirmed this Ansatz numerically for small numbers of
spins.
\end{thebibliography}
\end{document}